\documentclass[sigconf]{acmart}  

\AtBeginDocument{%
  \providecommand\BibTeX{{%
    \normalfont B\kern-0.5em{\scshape i\kern-0.25em b}\kern-0.8em\TeX}}}

\copyrightyear{2024}
\acmYear{2024}
\setcopyright{acmlicensed}\acmConference[CIKM '24]{Proceedings of the 33rd ACM International Conference on Information and Knowledge Management}{October 21--25, 2024}{Boise, ID, USA}
\acmBooktitle{Proceedings of the 33rd ACM International Conference on Information and Knowledge Management (CIKM '24), October 21--25, 2024, Boise, ID, USA}
\acmDOI{10.1145/3627673.3679638}
\acmISBN{979-8-4007-0436-9/24/10}

\usepackage{ulem}
\usepackage{subcaption}
\usepackage{ulem}
\usepackage{multirow}
\usepackage{diagbox}
\usepackage{graphicx}
\usepackage{algorithm, algpseudocode}
\usepackage{pifont}

\begin{document}

\author{Yunke Qu}
\affiliation{
  \institution{The University of Queensland}
  \city{Brisbane}
  \state{}
  \country{Australia}
}
\email{y.qu@uq.edu.au}

\author{Liang Qu}
\affiliation{
    \institution{Southern University of Science and Technology}
    \city{Shenzhen}
    \state{}
    \country{China}
}
\email{qul@mail.sustech.edu.cn}

\author{Tong Chen}
\affiliation{
  \institution{The University of Queensland}
  \city{Brisbane}
  \state{}
  \country{Australia}
}
\email{tong.chen@uq.edu.au}

\author{Xiangyu	Zhao}
\affiliation{
  \institution{City University of Hong Kong}
  \city{}
  \state{}
  \country{Hong Kong}
}
\email{xy.zhao@cityu.edu.hk}

\author{Quoc Viet Hung Nguyen}
\affiliation{
  \institution{Griffith University}
  \city{Gold Coast}
  \state{}
  \country{Australia}
}
\email{henry.nguyen@griffith.edu.au}

\author{Hongzhi Yin}
\authornote{Corresponding author}
\affiliation{
  \institution{The University of Queensland}
  \city{Brisbane}
  \state{}
  \country{Australia}
}
\email{h.yin1@uq.edu.au}

\renewcommand{\shortauthors}{Yunke Qu et al.}

\title{Scalable Dynamic Embedding Size Search for Streaming Recommendation} 

\begin{abstract}
Recommender systems typically represent users and items by learning their embeddings, which are usually set to uniform dimensions and dominate the model parameters. However, real-world recommender systems often operate in streaming recommendation scenarios, where the number of users and items continues to grow, leading to substantial storage resource consumption for these embeddings. Although a few methods attempt to mitigate this by employing embedding size search strategies to assign different embedding dimensions in streaming recommendations, they assume that the embedding size grows with the frequency of users/items, which eventually still exceeds the predefined memory budget over time. To address this issue, this paper proposes to learn Scalable Lightweight Embeddings for streaming recommendation, called SCALL, which can adaptively adjust the embedding sizes of users/items within a given memory budget over time. Specifically, we propose to sample embedding sizes from a probabilistic distribution, with the guarantee to meet any predefined memory budget. By fixing the memory budget, the proposed embedding size sampling strategy can increase and decrease the embedding sizes in accordance to the frequency of the corresponding users or items. Furthermore, we develop a reinforcement learning-based search paradigm that models each state with mean pooling to keep the length of the state vectors fixed, invariant to the changing number of users and items. As a result, the proposed method can provide embedding sizes to unseen users and items. Comprehensive empirical evaluations on two public datasets affirm the advantageous effectiveness of our proposed method. Code is available at https://github.com/qykcq/Scalable-Dynamic-Embedding-Size-Search-for-Streaming-Recommendation.
\end{abstract}



\maketitle

\section{Introduction}

Recommender Systems are algorithms designed to predict and suggest items or content that users might find appealing based on their preferences and behavior \cite{zhang2019deep, wang2021survey}. These systems play a crucial role in various applications, such as online streaming platforms, e-commerce websites, and social media, by enhancing user experience and engagement \cite{rececom1, rececom2, kywe2012survey}. However, a significant challenge in the implementation of recommender systems lies in the embedding tables, which are used to represent users and items in a low-dimensional embedding vectors of fixed and uniform sizes \cite{joglekar2020neural}. As the number of users and items grows in modern applications, the size of the embedding tables becomes a performance and memory bottleneck. For instance, considering Instagram with 2.4 billion monthly active users and 40 billion photos/videos, a conventional embedding table for the users and videos alone would require over 21,708 GB of memory. Managing and processing large embedding tables can lead to increased computational complexity, longer response times, and higher memory requirements, posing a challenge to the efficient operation of recommender systems in resource-constrained environments.

Many researchers have delved into mixed-dimension embedding size strategies aimed at minimizing memory usage while preserving the performance of recommender systems \cite{ciess, optembed, singleshot, liang2023lcc, chentongkdd2021, bet, liu2021learnable, joglekar2020neural}. These strategies involve assigning different embedding sizes based on the characteristics of users and items, such as their interaction counts. However, these methods are tailored for static and non-streaming settings and the embedding sizes remain fixed after allotment, which largely limits their practical utility in streaming settings for two reasons. Firstly, in streaming environments, where users and items exhibit high dynamism, the embedding sizes need to be continually adjusted to accommodate the evolving user and item behavior. For example, the embedding size of a newly released movie may need to be increased as its frequency rises over time. Maintaining consistent embedding sizes in such a situation will lead to performance degradation because it fails to adopt to the changing frequency levels. Secondly, these methods cannot to assign embedding sizes to newly joined users or items. As a result, implementing these methods in streaming scenarios necessitates retraining them from scratch at the onset of each time segment, introducing significant computational overheads.

In this regard, several embedding size search mechanisms have emerged to address the challenges of adaptability and efficiency in the evolving landscape of recommendation applications, particularly in streaming scenarios. For instance, Liu et al. \cite{autoemb} proposed AutoEmb, a framework treating the embedding size search as a multi-class classification problem and utilizing a neural network for soft selection of embedding sizes for users and items. The final output embedding size may vary according to the input frequency. Similarly, ESAPN \cite{esapn} and DESS \cite{dess} formulate the problem of embedding size adjustment as a binary classification problem. They employ external policies to determine whether to enlarge the embedding size for each user and item as data streams in. However, all of these frameworks lack precise control over their parameter usage, making it challenging to set predefined memory budgets that is crucial for resource-constrained applications. Although BET \cite{bet} partially addresses this in static recommendation systems using table-level search, which computes the embedding sizes for all users and items, it falls short in streaming settings due to its inability to allocate embedding sizes to newly introduced users and items. 

Another problem with ESAPN and DESS is their limited flexibility in adjusting the embedding sizes of previous users and items. Unlike AutoEmb, these methods can only increase embedding sizes for users and items that become more popular, but they cannot reduce them when frequency wanes. Assigning large embedding sizes to users or items that were only popular in the past is unnecessary and inefficient in terms of memory usage.  In summary, it is essential to develop a method for embedding size search in streaming recommender systems that have the following advantages: (1) \textbf{Dynamic Sizes}: it can automatically adjust (increase or decrease) embedding sizes to adopt to the changing frequency in each time segment without the need to retrain, (2) \textbf{Controllable Budget}: it can easily scale the embedding table to meet any predefined parameter budget without extra hyperparameter tuning, and (3) \textbf{Streaming Friendly}: it can be easily implemented in streaming settings without retraining in each time segment.

To resolve the three challenges, we propose \textbf{Scal}able \textbf{L}ightweight Embeddings (SCALL) for streaming recommendation systems, an automated embedding size search policy based on the reinforcement learning algorithm Soft Actor-Critic (SAC) \cite{sac}. SCALL achieves (1) and (3) by leveraging the reservoir sampling mechanism of SAC, which enables continual policy learning without the need for periodic retraining. SCALL samples embedding sizes from a probabilistic distribution whose parameters are predicted by the policy network. A controllable memory budget can be effortlessly enforced during the embedding size sampling process, thus achieving (2). Table \ref{tab:comparison} provides a brief comparison between our proposed method and two other state-of-the-art embedding size search methods, showcasing the superiority of SCALL. 

\begin{table}[htbp]
\caption{Comparison between our method and the most representative state-of-the-art embedding size search methods.
}
\resizebox{0.9\columnwidth}{!}{%
\begin{tabular}{cccccc}
\toprule
Benefits & CIESS & BET & \begin{tabular}[c]{@{}c@{}}ESAPN\\ DESS\end{tabular} & AutoEmb & SCALL \\ \hline
Dynamic Sizes & \ding{56} & \ding{56} & \ding{56} & \ding{52} & \ding{52}  \\ \midrule
Controllable Budget & \ding{56} & \ding{52} & \ding{56} & \ding{56} & \ding{52}  \\ \midrule
Streaming Friendly & \ding{56} & \ding{56} & \ding{52} & \ding{52} & \ding{52}  \\ \midrule
\end{tabular}%
}
\label{tab:comparison}
\end{table}

Since more frequent users and items usually require higher embedding dimensions, incorporating frequency information into the learning of optimal embedding sizes has been highlighted in several studies \cite{autoemb, esapn, ciess, bet}. However, as the number of users and items increases over time in streaming settings, the size of the state vector becomes variable. To address this issue, we propose the mean pooling strategy, which involves sorting and grouping users and items based on their frequency. Subsequently, we include only the average frequency of each group in each state, ensuring that the length of the state vector remains fixed regardless of the varying number of users and items. Additionally, by sampling embedding sizes from a probabilistic distribution, our proposed method achieves independence from the fluctuating number of users and items, thereby fulfilling (3). The contributions of this paper are outlined below:
\begin{itemize}
    \item We pinpoint the practical bottlenecks of current embedding size search methods for streaming recommender systems, specifically their inability to adaptively adjust embedding sizes to conform to a predefined parameter budget.
    \item We introduce SCALL, an innovative reinforcement learning (RL)-based approach that dynamically allocates parameters to all the users and items with the guarantee to meet any predefined memory budget in streaming settings, regardless of the changing number of users and items.
    \item We improve the table-level search paradigm to accommodate the varying number of users and items in streaming recommendation systems and use it to improve search efficiency.
    \item We conduct comprehensive empirical evaluations on two public datasets and compare SCALL to various embedding sparsification algorithms, affirming the advantageous efficacy of SCALL.
\end{itemize}

\section{Relative Work}
Before delving into our proposed method, we present the pertinent research topics to offer an in-depth research context from three perspectives: deep recommender systems, streaming recommender systems, and embedding size search in recommendation.

\subsection{Deep Recommender Systems}
Researchers have put forth a variety of deep recommender models aimed at capturing user-item relations. The initial wave of deep recommender systems is rooted in Multi-Layer Perceptrons (MLPs) \cite{cheng2016wide, he2017neural}. Cheng et al. \cite{cheng2016wide} introduced Wide\&Deep, a model that combines a wide linear model with a deep neural network by computing the weighted sum of their output to leverage the benefits of both memorization and generalization. Another example is Neural Collaborative Filtering \cite{he2017neural}, which integrates an MLP with matrix factorization to grasp the nuanced user-item interactions. Beyond MLP-based methods, researchers have explored graph-based architectures \cite{wang2019neural, he2020lightgcn, graphaug}. For instance, Neural Graph Collaborative Filtering \cite{wang2019neural} expands neural collaborative filtering by incorporating graph convolution networks to model user-item interactions. LightGCN \cite{he2020lightgcn} simplifies NGCF by eliminating self-connections, feature transformations, and linear activation operations. UltraGCN \cite{ultragcn} further improves LightGCN by using a constraint loss to approximate the limit of infinite-layer graph convolutions. In addition to models that learn from user-item interactions, other data modalities such as location \cite{yin2015jmo, wang2020npo}, sentiment \cite{wang2017als}, activity level \cite{changshuozhang2024rlt}, text \cite{jiachengli2023tia, changshuozhang2024qagcf}, images \cite{suhangwang2017www}, and videos \cite{lee2018collaborative} have also been exploited for recommendation.

\subsection{Streaming Recommender Systems}
 To adapt to ever-evolving landscape of recommendation applications, online learning algorithms that update recommendation model as new data streams in have been studied for streaming recommender systems. For instance, Wang et al. \cite{SPMF} provided a Stream-centered Probabilistic Matrix Factorization model. Guo et al. \cite{sssreckdd2019} proposed a Streaming Session-based Recommendation Machine to tackle the challenges introduced by user behaviors and large session data volume in streaming recommendation. Qiu et al. \cite{gag} introduced a novel Global Attributed Graph neural network model, which learns better session and user representations. Wang et al. \cite{qinyong2018kdd} proposed a streaming recommender model to capture long-term stable and short-term interests. It is further enhanced by an adaptive negative sampling framework based on Generative Adversarial Nets. To mitigate the problem of catastrophic forgetting and preserve a sketch of users' long-term interests, reservoir sampling techniques have also been proposed \cite{tsample, chang2017www, terec, gag}. They typically involve storing historical user-item interactions in a reservoir, where interactions are sampled using various strategies to update the recommender model. Although these methods can learn user and item representations from continuous streaming data, they still face memory efficiency issues which are yet to be addressed.

\subsection{Embedding Compression in Recommendation}
A variety of research endeavors have investigated methods for learning compressed embeddings, serving as a memory-efficient alternative to conventional embedding tables in recommender systems, as outlined in \cite{automlrecsurvey, yin2024ondevicerec, tran2024thoroug}. Approaches like feature selection search, as exemplified by \cite{autofield, luo2019autocross, adafs}, focus on training models to learn importance scores for effective filtering of unimportant feature fields. Embedding dimension search methods, illustrated by \cite{liu2021learnable, singleshot, autosrh, ciess, bet}, enhance traditional recommender systems by introducing mixed dimensions for features, departing from uniform sizes. Compositional embedding methods \cite{liang2024legcf, liang2023lcc} represents users and items with a smaller number of embedding vectors (aka. meta-embeddings) to shrink the embedding size. Elastic embedding \cite{chentongkdd2021, zhengruiqi2024pee} enable users and items to share embedding blocks. The knowledge distillation method \cite{xiaxin2022odn, xiaxin2023eod} compresses a model by having a smaller student model learn from a larger teacher model. SCALL maps each user and item to their optimal embedding size so it belong to category of embedding dimension search methods.

Most of the previously mentioned methods offer sparsified embeddings in static recommendation settings. In the context of streaming recommendation tasks, \cite{dess, esapn, autoemb} dynamically assign embedding sizes to users and items as data streams in. SCALL is also part of this category, yet it stands out as the first method that combines the three aforementioned advantages: dynamic sizes, controllable budget and streaming friendly.

\section{Problem Definition}
In this section, we define the task of streaming recommendation and dynamic embedding size prediction.

We create a new data segments each time we witness $m$ new users and items as the data streams in. As a result, the entire data stream is partitioned into $T$ consecutive and mutually exclusive segments arranged in chronological order: $[D_1, \cdots, D_{T-1}, D_{T}]$. Subsequently, each segment $D_t \in \{D_1, \cdots, D_{T-1}, D_{T}\}$ is further bifurcated into two exclusive parts: the training portion $D_{t}^{tr}$ and the test portion $D_{t}^{te}$. Building upon this framework, the streaming recommendation task is framed as follows: given $[D_1^{tr}, \cdots, D_{T-1}^{tr}, D_T^{tr}]$, the objective is to train a model $G(\cdot)$ to predict the user preferences on $[D_1^{te}, \cdots, D_{T-1}^{te}, D_T^{te}]$. In each time segment $t$, we also build a reservoir, denoted as $\mathcal{R}_t$, containing historical interactions that are randomly sampled from $D_1 \cup ... \cup D_{t-2} \cup D_{t-1}^{tr}$ to mitigate catastrophic forgetting. The reservoir sampling strategy is the same as in \cite{sssreckdd2019}. 

SCALL operates in a streaming setting. It starts by initializing a base recommender $G_0(\cdot)$. The algorithm then iterates through each time segment of the data stream. In each time segment $t \in [1, T]$ with $(D_t^{tr} \cup D_t^{te})$, the embedding size predictor $F(\cdot)$ is first trained on $D^{tr}_{t-1}$. After the training is completed, $F(\cdot)$ should provide the embedding size for each user and item in $D_1 \cup ... \cup D_{t-1} \cup D_t^{tr}$. After obtaining the embedding sizes from the embedding size predictor, the base recommender $G_{t-1}(\cdot)$ should adjust its embedding sizes and update itself to $G_{t}(\cdot)$ on the training samples $D^{tr}_t$. After this update, we evaluate $G_{t}(\cdot)$ on $D^{te}_t$ and report the recommendation quality w.r.t. Recall@20 and NDCG@20. The algorithm alternates between updating the embedding size predictor and the base recommender in a coordinated manner. The goal of embedding size search is to provide embedding sizes to maximize the recommendation quality on $D^{te}_t$ for each time segment $t$. In the remainder of this paper, we omit the subscript $t$ and denote the base recommender as $G(\cdot)$ when there is no confusion. 

\section{Proposed Methodology}

\begin{figure}
    \centering
    \includegraphics[width=0.9\linewidth]{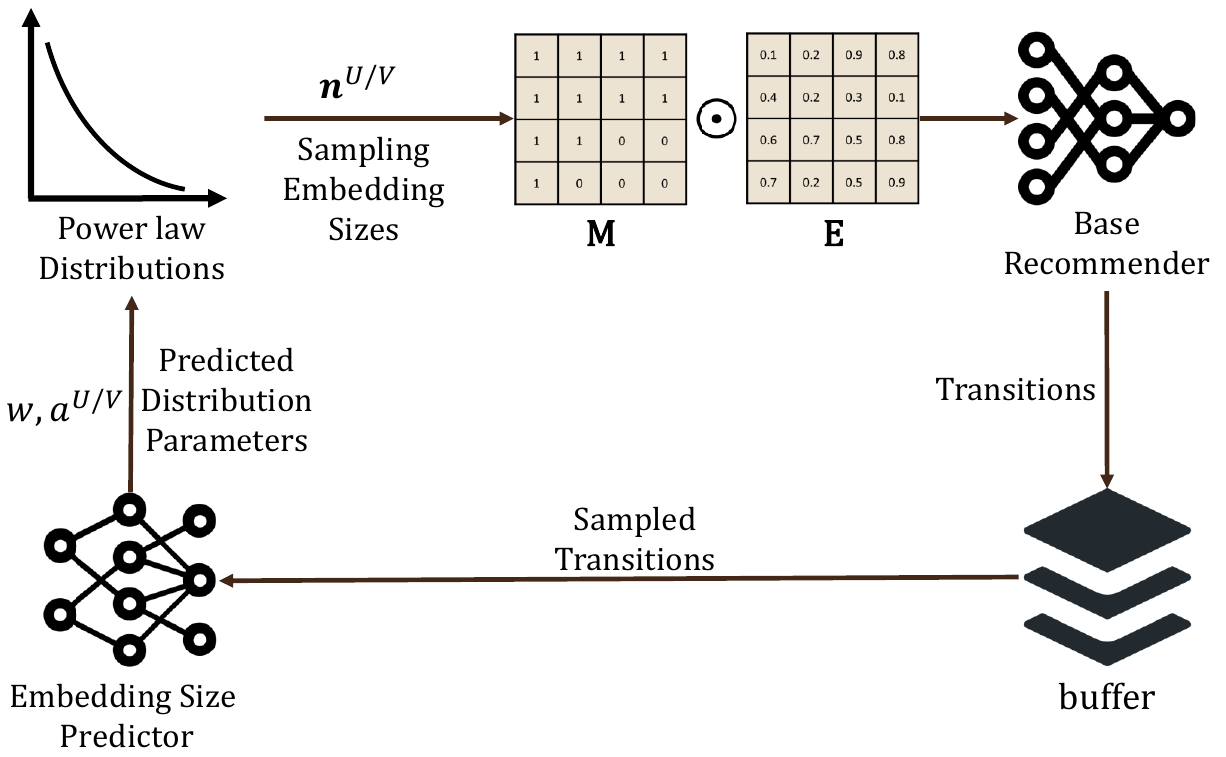} 
    \caption{An overview of SCALL.} \label{fig:overview}
\end{figure}

In a nutshell, SCALL comprises two key components that work alternately: a pre-trained base recommender $G(\cdot)$ parameterized by $\Theta$, and a reinforcement learning-based embedding size predictor $F(\cdot)$ parameterized by $\Phi$. Figure~\ref{fig:overview} illustrates the workflow of SCALL. In each time segment $t \in [1, T]$, $F(\cdot)$ computes actions that determine the portion of parameters allocated to the users and the parameters of the distributions from which we sample embedding sizes. A fresh base recommender is initialized and adjusts its user/item embedding sizes and tunes its parameters using the training samples from the previous segment $D_{t-1}^{tr}$ with early stopping. Then the base recommender is evaluated on a held-out test set $D_{t-1}^{te}$ w.r.t. Recall@20 and NDCG@20. Based on  early stopping Recall@20 and NDCG@20 scores, the embedding size predictor $F(\cdot)$ is revised, and thus the embedding size selection is updated for subsequent iterations. The following sections elucidate the design of SCALL.

\subsection{Base Recommender with Adjustable Embedding Sizes} \label{subsec:baserec}
We denote the set of all users as $\mathcal{U}$ and the set of all items $\mathcal{V}$. The recommender's embedding table maps each user or item to embedding vectors $\mathbf{e}_{n} \in \mathbb{R}^{d_{max}}$, where $n\in\mathcal{U}\cup \mathcal{V}$ and $d_{max}$ is the full embedding size. The vertical concatenation of the embedding vectors for all users and items results in the embedding table $\mathbf{E} \in \mathbb{R}^{(|\mathcal{U}|+|\mathcal{V}|) \times d_{max}}$. Following the common practice for embedding sparsification in \cite{liu2021learnable, optembed, ciess, bet}, we adjust the embedding sizes through a binary mask matrix $\mathbf{M}\in \{0,1\}^{(|\mathcal{U}|+|\mathcal{V}|) \times d_{max}}$, where each row $\mathbf{m}_n \in \mathbf{M}$ is a binary mask vector for user/item $n$, where a proposed embedding size $d_n$ sets the first $d_n$ elements to ones and the subsequent $d_{max} - d_n$ elements to zeros:
\begin{equation}
    \mathbf{m}_n[s] = 
    \begin{cases}
        1 & \text{for }1 \le s \le d_n\\    
        0 & \text{for }d_n < s < d_{max}
    \end{cases}, \;\;\;\; n\in\mathcal{U}\cup \mathcal{V}.
\end{equation}
The embedding table is adjusted during look-up via element-wise multiplication with the binary mask matrix $\mathbf{M}$, i.e., $\mathbf{e}_{n} = (\mathbf{E} \odot \mathbf{M})_n$. $\mathbf{M}$ is iteratively updated based on actions given by the policy network. During deployment, only the heavily sparsified embedding table $\mathbf{E}_{sparse} = \mathbf{E} \odot \mathbf{M}$ is stored, where sparse matrix storage techniques \cite{virtanen2020scipy, sedaghati2015automatic} handle zero-valued entries efficiently. The recommender model $G(\cdot)$ computes the pairwise user-item similarity $\hat{y}_{uv}$ using sparse embedding vectors:
\begin{equation}
    \hat{y}_{uv} = G(\mathbf{e}_u,\mathbf{e}_{v}), \;\;\;\; u\in\mathcal{U}, v\in \mathcal{V}.
\end{equation}

For optimization, the Bayesian Personalized Ranking loss of the base recommender $G(\cdot)$ on training samples $D^{tr}$ \cite{rendle_bpr_2012} is adopted:
\begin{equation} \label{eq:bpr}
   \mathcal{L}_{BPR}(G(\cdot)|D^{tr}) = \sum_{(u, v, v') \in \mathcal{D}^{tr}} - \ln \sigma (\hat{y}_{uv} - \hat{y}_{uv'}) + \eta\|\Theta\|_2^2,
\end{equation}
where $(u, v, v')$ denotes user $u$'s preference for item $v$ to $v'$, $\Theta$ is the parameter of $G(\cdot)$, and $\eta\|\Theta\|_2^2$ provides regularization. The goal is to each user and item to their optimal embedding size under a strict memory budget with a fixed mean embedding size: or with a fixed number of parameters, expressed in the overall objective:
\begin{equation}\label{eq:objective1}
    \min_{\Theta,\Phi} \mathcal{L}_{BPR} \;\;\;\text{s.t.}\; \frac{\| \mathbf{M}\|_{1,1}}{(|\mathcal{U}|+|\mathcal{V}|)} \leq c 
\end{equation}
where $\| \mathbf{M}\|_{1,1}$ represents the number of ones in the binary mask, $c$ is the desired mean embedding size over all users and items, and $B$ is the desired number of parameters. Alternatively, we can also fix the total number of parameters to $B$:
\begin{equation}
    \min_{\Theta,\Phi} \mathcal{L}_{BPR} \;\;\;\text{s.t.}\;\| \mathbf{M}\|_{1,1} \leq B
\end{equation}
With fixed $c$, the total number of parameters can be calculated as follows:
\begin{equation}\label{eq:objective2}
    B = c(|\mathcal{U}|+|\mathcal{V}|)
\end{equation}

\subsection{Mixed Embedding Size Search with Reinforcement Learning}

\subsubsection{Environment} The base recommender is the environment. It returns the top-$k$ items for each user given their embedding sizes.

\begin{figure}
    \centering
    \includegraphics[width=0.9\linewidth]{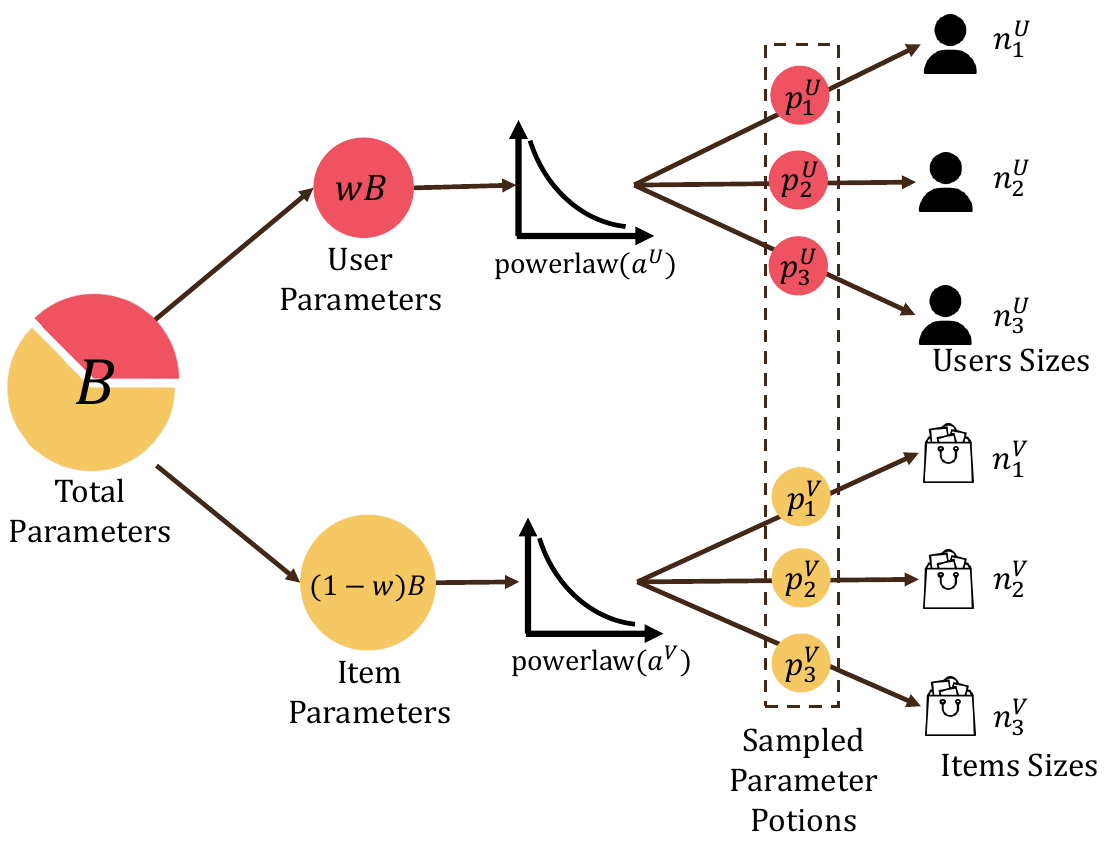} 
    \caption{The total parameters are allocated to to 3 users and 3 items in this example. The size of each circle corresponds to the number of parameters, with a larger circle indicating a higher parameter count.} \label{fig:allocation}
\end{figure}

\subsubsection{Action} \label{subsec:action} We allocate $wB$ parameters to the users and $(1-w)B$ to the items, where $w \in (0, 1)$ denotes the portion of parameters assigned to the users. The users are sorted in descending order of their frequency in $D^{tr}_{t} \cup \mathcal{R}_t$. If two users or items have equal frequency, preference is given to the one that appears earlier in the sorted list. Let $\boldsymbol{p}^U$ and $\boldsymbol{p}^V$ represent the percentage of parameters assigned to each user and item. Inspired by the use of Power Law distribution to model income distribution, we adopt two Power Law distributions, parameterized by $\alpha^U, \alpha^V \in (\alpha_{min}, \alpha_{max})$ to separately sample values for $\boldsymbol{p}^U$ and $\boldsymbol{p}^V$. We can adjust the fairness of parameter allocation by adjusting the values of $\alpha^U$ and $\alpha^V$. The parameters are uniformly shared by each user/item when $\alpha^U$ and $\alpha_V$ approach to $\alpha_{max}$; when $\alpha^U$ and $\alpha_V$ approach to $\alpha_{min}$, the parameters tend to be assigned to the most frequent user/items. In other words, parameter assignment to each user and item is modeled by a mixture model, where each component is a Power Law distribution: $\text{powerlaw}(\alpha) = \alpha x^{\alpha - 1}$:
\begin{align} \label{eq:sampling}
    \boldsymbol{q}^{U} \sim \text{powerlaw}(\alpha^{U}) \nonumber ;\;\;\;
    \boldsymbol{q}^{V} \sim \text{powerlaw}(\alpha^{V})
\end{align}
Subsequently, the distributions are normalized to ensure that the sum of elements in each distribution is equal to 1:
\begin{equation} \label{eq:normalization}
    \boldsymbol{p}^{U} = \frac{\boldsymbol{q}^{U}}{\sum_{u \in |\mathcal{U}|} \boldsymbol{q}_u^{U}} \nonumber ;\;\;\;
    \boldsymbol{p}^{V} = \frac{\boldsymbol{q}^{V}}{\sum_{v \in |\mathcal{V}|}\boldsymbol{q}_v^{V}}
\end{equation}
Notably, we separate the users and items when sampling their embedding sizes to accommodate the distributional difference in the user and item frequencies. Given a fixed total parameter count, the embedding sizes $\boldsymbol{n}^{U}$ for each user and $\boldsymbol{n}^{V}$ for each item can be easily calculated as follows:
\begin{equation}
    \boldsymbol{n}^{U} = \lfloor wB\boldsymbol{p}^U \rfloor \nonumber ;\;\;\;
    \boldsymbol{n}^{V} = \lfloor (1-w)B\boldsymbol{p}^V \rfloor
\end{equation}

Figure \ref{fig:allocation} illustrates this parameter allocation process. Clearly, once we decide the values for $w$, $\alpha^U$ and $\alpha^V$, we can easily derive the embedding size for every user and item. Therefore, we formulate each action as:
\begin{equation}
    a = (w, \alpha^U, \alpha^V)
\end{equation}

\subsubsection{State} \label{subsec:state} The input state $s$ is vital in the decision-making process of the policy network by capturing relevant information. Prior research, such as \cite{autoemb, esapn}, has highlighted the effectiveness of integrating factors like frequency and contextual details like reward. Following the same design, we integrate the previous reward $r$ in the state. 

\begin{figure}
    \centering
    \includegraphics[width=0.75\linewidth]{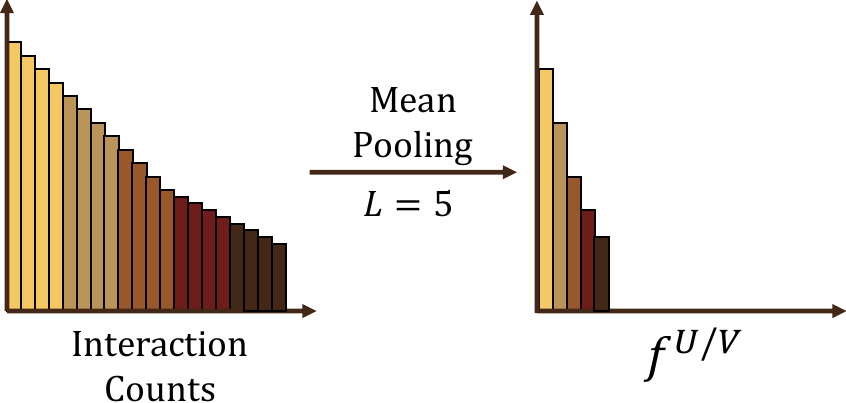} 
    \caption{Mean pooling is used to achieve fixed-length vectors incorporating frequency information. The users or items in this example are divided into 5 groups.} \label{fig:meanpooling}
\end{figure}

However, we cannot directly incorporate the frequency of every user and item in the state since the number of users and items varies in each data segment. To keep the length of the state fixed, after normalizing the user interaction counts to the range of (0, 1) using minmax normalization, we sort the users in descending order of their frequency and divide them into $L$ groups. Then we calculate the mean frequency of each group and include it in the state. This practice is inspired by the mean pooling operation in Convolution Neural Networks. Figure \ref{fig:meanpooling} provides an illustration of this process. The same operation is conducted on the item interaction counts. The resulting pooled user and item frequency is denoted as $\boldsymbol{f}^U$ and $\boldsymbol{f}^V$.

Likewise, the resulting embedding sizes $\boldsymbol{n}$ cannot be included in the state as contextual information due to their varying length. In light of this challenge, we only include the range of $\boldsymbol{n}^U$ and $\boldsymbol{n}^V$, normalized by $d_{max} - d_{min}$, to measure their dispersion:
\begin{equation}
    h^{U} = \frac{\text{max}(\boldsymbol{n}^U) - \text{min} (\boldsymbol{n}^U)}{d_{max} - d_{min}}  \nonumber ;\;\;\;
    h^{V} = \frac{\text{max}(\boldsymbol{n}^V) - \text{min}(\boldsymbol{n}^V)}{d_{max} - d_{min}}
\end{equation}
It is worth noting that variance or standard deviation cannot be used here because they are sensitive to the scale of the data. Although $\alpha^U$ or $\alpha^V$ can measure the statistical dispersion of $\boldsymbol{n}$, they are not suitable for use here because they are invariant of the data scale, which can be exemplified by the case of $\boldsymbol{n}_1 = [128, 64, 64]$,  $\boldsymbol{n}_2 = [2, 1, 1]$ $\boldsymbol{n}_3 = [1, 1, 1]$. With the use of $\alpha^{U/V}$, $\boldsymbol{n}_1$ and $\boldsymbol{n}_2$ will be deemed as equivalent, while in reality $\boldsymbol{n}_2$ makes little difference from $\boldsymbol{n}_3$ in affecting the performance of the base recommender when they are used as embedding sizes.

Moreover, to capture the characteristics of the data segment, the mean user/item embeddings (denoted as $\boldsymbol{m}^U, \boldsymbol{m}^V$) and the user number to item number ratio $\frac{|\mathcal{U}|}{|\mathcal{U}|+|\mathcal{V}|}$ are also included by the state. The mean embeddings are calculated as follows:
\begin{equation}
    \boldsymbol{m}^U = \frac{\sum_{u \in \mathcal{U}}(\mathbf{E} \odot \mathbf{M})_u}{|\mathcal{U}|} \nonumber ;\;\;\;
    \boldsymbol{m}^V = \frac{\sum_{v \in \mathcal{V}}(\mathbf{E} \odot \mathbf{M})_v}{|\mathcal{V}|}
\end{equation}
Similar to the computation of $\boldsymbol{f}^{U}$ and $\boldsymbol{f}^{V}$, we take the mean of the user and item embedding sizes because their vector length is independent of the number of users and items.

In summary, our method constructs the state $s$ as follows: 
\begin{equation}\label{eq:state}
    s = (\boldsymbol{f}^{U}, \boldsymbol{f}^{V}, r, \frac{|\mathcal{U}|}{|\mathcal{U}|+|\mathcal{V}|}, h^U, h^V, \boldsymbol{m}^U, \boldsymbol{m}^V)
\end{equation}

\subsubsection{Reward} In each iteration, the base recommender dynamically adjusts its embedding sizes based on the action $a$. Subsequently, a reward $r$ is generated to provide feedback to the current policy and guide the following embedding size allocations. Aligned with our objective equations Eq.(\ref{eq:objective1}) and Eq. (\ref{eq:objective2}), the reward needs to effectively capture the recommendation quality. Building on the findings in \cite{white2021how}, which highlight the effectiveness of accuracy with early stopping as a performance predictor in neural architecture search, we utilize Recall@20 and NDCG@20 with early stopping to formulate the reward. Specifically, a fresh base recommender is initialized for fine-tuning before its recommendation quality is evaluated. We initialize a new recommender and do not inherit any pre-trained embeddings so that the previous embedding sizes do not have any influence on the ongoing embedding size search. The agent will have the tendency to preserve the previous embedding sizes if we directly fine-tune a trained base recommender. Hence, we define the following reward for an action $a$ in time segment $t$:
\begin{equation} \label{eq:reward}
    r  \propto \frac{\text{eval}(G'_{t,j}| \mathcal{D}^{te}_t)}{\text{eval}(G_{t,j}| \mathcal{D}^{te}_t)},
\end{equation}
where $r$ is further normalized to fall within the range of $(0, 10)$, $G_{t,j}$ represents the snapshot of the base recommender that is updated on $D^{tr}_t$ at the $j$-th step, $G'_{t,j}$ represents the newly initialized base recommender after being fine-tuned on $D^{tr}_t$ for $j$ steps, $\text{eval}(\cdot|\mathcal{D}_t^{te})$ is a recommendation quality measure on the test set $\mathcal{D}_t^{te}$. It is constructed as an ensemble of Recall@20 and NDCG@20, which are both common recommendation metrics \cite{he2020lightgcn,rendle_bpr_2012}: 
\begin{equation}\label{eq:eval}
    \text{eval}(\cdot) = \sum_{u \in \mathcal{U}} \frac{\text{Recall@}20_u + \text{NDCG@}20_u}{2|\mathcal{U}|}
\end{equation}
$G_{1,j}$ is obtained by training a freshly initialized recommender on $D_1^{tr}$ with fixed and uniform sizes.

Importantly, Recall@20 and NDCG@20 scores are not directly comparable across different time segments due to the inherent variability in data characteristics. Therefore, instead of using $\text{eval}(G'_{t,j}|\mathcal{D}^{te}_t)$ as the reward, we compare it to $\text{eval}(G_{t,j}|\mathcal{D}^{te}_{t})$ to reflect the performance gain/loss. This scaling ensures comparability across each time segment, allowing it to reflect performance fluctuations with respect to the performance before the embedding size adjustment.

\subsubsection{Actor and Critic} SCALL adopts the actor-critic paradigm in reinforcement learning. We employ Soft Actor-Critic (SAC) \cite{sac} as the reinforcement learning backbone. SAC consists of an actor network $\pi(\cdot)$ and a critic network $Q(\cdot)$. In each iteration, the actor network $\pi(\cdot)$ maps current state $s$ to an action $a$:
\begin{align}\label{eq:actor}
    a = \pi(s) 
\end{align}

To approximate the quality (Q-value) of an action $a$ taken at state $s$, we employ a critic network denoted as $Q(\cdot)$. The critic networks are updated using transition tuples stored in the replay buffer $\mathcal{Z}$. The complete pseudo-code of SCALL is presented in Algorithm \ref{alg:cap}.

\begin{algorithm}[t]
\caption{SCALL}
\label{alg:cap}
\begin{algorithmic}[1]
\State Initialize replay buffer $\mathcal{Z}$;
\State Initialize base recommender $G_0(\cdot)$;
\For{$t = 1, \cdots, T$}
    \If{t > 1}
        \For{$i = 0,\cdots,N$}
            \State /*\; Update embedding size predictor \;*/
            \State Get last time segment of data $(D^{tr}_{t-1} \cup D^{te}_{t-1})$;
            \State Obtain $a_i \leftarrow$ Eq.(\ref{eq:actor}) and temporarily update $\mathbf{M}$;    
            \State Tune the newly initialized $G'_{t-1}(\cdot)$ on $D^{tr}_{t-1}$;
            \State Calculate reward $r_i \!\!\leftarrow\!$ Eq.(\ref{eq:reward}) 
            \State Get subsequent state $s’ \!\!\leftarrow\!$ Eq.(\ref{eq:state});
            \State Update buffer $\mathcal{Z} \leftarrow \mathcal{Z}\cup (s_i, a_i, r_i, s'_i)$; 
            \State Randomly sample a batch of transitions from $\mathcal{Z}$;
            \State Update $Q(\cdot)$ and $\!\pi(\cdot)$ as in SAC \cite{sac};
        \EndFor 
    \EndIf
    \State /*\; Update base recommender \;*/
    \State Get current time segment of data $(D^{tr}_{t} \cup D^{te}_{t})$;
    \If{t = 1}
        \State Update $\mathbf{M}$ with fixed and uniform sizes;
    \Else:
        \State Obtain action $a \leftarrow$ Eq.(\ref{eq:actor}) and permanently update $\mathbf{M}$;
    \EndIf
    \State Permanently update $G_{t-1}(\cdot)$ to $G_t(\cdot)$ on $D^{tr}_{t}$;
    \State Evaluate $G_t(\cdot)$ on $D^{te}_{t}$ w.r.t. Recall@20 and NDCG@20;
\EndFor
\end{algorithmic}
\end{algorithm}

\begin{table*}[]
\caption{Performance of all methods on Yelp and Amazon-Book  with mean embedding size set to 64, 32, and 16. The best results are highlighted. Recall@20 and NDCG@20 are abbreviated to R@20 and N@20 in this and all the subsequent tables for simplicity. All our results are statistically significant with $p < 0.02$.}
\resizebox{0.9\textwidth}{!}{%
\begin{tabular}{|c|cccccc|cccccc|}
\hline
 & \multicolumn{6}{c|}{Yelp} & \multicolumn{6}{c|}{Amazon-Book} \\ \hline
\multirow{2}{*}{Method} & \multicolumn{2}{c|}{c = 16} & \multicolumn{2}{c|}{c = 32} & \multicolumn{2}{c|}{c = 64} & \multicolumn{2}{c|}{c = 16} & \multicolumn{2}{c|}{c = 32} & \multicolumn{2}{c|}{c = 64} \\ \cline{2-13} 
 & \multicolumn{1}{c|}{R@20} & \multicolumn{1}{c|}{N@20} & \multicolumn{1}{c|}{R@20} & \multicolumn{1}{c|}{N@20} & \multicolumn{1}{c|}{R@20} & N@20 & \multicolumn{1}{c|}{R@20} & \multicolumn{1}{c|}{N@20} & \multicolumn{1}{c|}{R@20} & \multicolumn{1}{c|}{N@20} & \multicolumn{1}{c|}{R@20} & N@20 \\ \hline
ES & \multicolumn{1}{c|}{0.0468} & \multicolumn{1}{c|}{0.0253} & \multicolumn{1}{c|}{0.0524} & \multicolumn{1}{c|}{0.0281} & \multicolumn{1}{c|}{0.0571} & 0.0303 & \multicolumn{1}{c|}{0.0411} & \multicolumn{1}{c|}{0.0229} & \multicolumn{1}{c|}{0.0478} & \multicolumn{1}{c|}{0.0265} & \multicolumn{1}{c|}{0.0497} & 0.0279 \\ \hline
MR & \multicolumn{1}{c|}{0.0453} & \multicolumn{1}{c|}{0.0243} & \multicolumn{1}{c|}{0.0488} & \multicolumn{1}{c|}{0.0260} & \multicolumn{1}{c|}{0.0526} & 0.0281 & \multicolumn{1}{c|}{0.0376} & \multicolumn{1}{c|}{0.0208} & \multicolumn{1}{c|}{0.0416} & \multicolumn{1}{c|}{0.0229} & \multicolumn{1}{c|}{0.0452} & 0.0249 \\ \hline
CIESS & \multicolumn{1}{c|}{0.0536} & \multicolumn{1}{c|}{0.0295} & \multicolumn{1}{c|}{0.0564} & \multicolumn{1}{c|}{0.0309} & \multicolumn{1}{c|}{0.0585} & 0.0319 & \multicolumn{1}{c|}{0.0419} & \multicolumn{1}{c|}{0.0235} & \multicolumn{1}{c|}{0.0484} & \multicolumn{1}{c|}{0.0274} & \multicolumn{1}{c|}{0.0520} & 0.0293 \\ \hline
BET & \multicolumn{1}{c|}{0.0542} & \multicolumn{1}{c|}{0.0297} & \multicolumn{1}{c|}{0.0581} & \multicolumn{1}{c|}{0.0318} & \multicolumn{1}{c|}{0.0596} & 0.0324 & \multicolumn{1}{c|}{0.0431} & \multicolumn{1}{c|}{0.0243} & \multicolumn{1}{c|}{0.0501} & \multicolumn{1}{c|}{0.0282} & \multicolumn{1}{c|}{0.0529} & 0.0300 \\ \hline
SCALL & \multicolumn{1}{c|}{\textbf{0.0560}} & \multicolumn{1}{c|}{\textbf{0.0305}} & \multicolumn{1}{c|}{\textbf{0.0604}} & \multicolumn{1}{c|}{\textbf{0.0329}} & \multicolumn{1}{c|}{\textbf{0.0623}} & \textbf{0.0335} & \multicolumn{1}{c|}{\textbf{0.0461}} & \multicolumn{1}{c|}{\textbf{0.0258}} & \multicolumn{1}{c|}{\textbf{0.0516}} & \multicolumn{1}{c|}{\textbf{0.0290}} & \multicolumn{1}{c|}{\textbf{0.0554}} & \textbf{0.0310} \\ \hline
 & \multicolumn{2}{c|}{c} & \multicolumn{2}{c|}{R@20} & \multicolumn{2}{c|}{N@20} & \multicolumn{2}{c|}{c} & \multicolumn{2}{c|}{R@20} & \multicolumn{2}{c|}{N@20} \\ \hline
ESAPN & \multicolumn{2}{c|}{84} & \multicolumn{2}{c|}{0.0592} & \multicolumn{2}{c|}{0.0323} & \multicolumn{2}{c|}{78} & \multicolumn{2}{c|}{0.0529} & \multicolumn{2}{c|}{0.0299} \\ \hline
DESS & \multicolumn{2}{c|}{39} & \multicolumn{2}{c|}{0.0586} & \multicolumn{2}{c|}{0.0318} & \multicolumn{2}{c|}{46} & \multicolumn{2}{c|}{0.0503} & \multicolumn{2}{c|}{0.0285} \\ \hline
AutoEmb & \multicolumn{2}{c|}{74} & \multicolumn{2}{c|}{0.0560} & \multicolumn{2}{c|}{0.0307} & \multicolumn{2}{c|}{65} & \multicolumn{2}{c|}{0.0489} & \multicolumn{2}{c|}{0.0277} \\ \hline
\end{tabular}%
}
\label{tab:fixedc}
\end{table*}

\begin{table*}[]
\caption{Performance of all methods on Yelp and Amazon-Book with the parameter budget set to 10MB and 5MB. The best results are highlighted. All our results are statistically significant with $p < 0.03$.}
\resizebox{0.63\textwidth}{!}{%
\begin{tabular}{|c|cccc|cccc|}
\hline
 & \multicolumn{4}{c|}{Yelp} & \multicolumn{4}{c|}{Amazon-Book} \\ \hline
\multirow{2}{*}{Method} & \multicolumn{2}{c|}{B = 5MB} & \multicolumn{2}{c|}{B = 10MB} & \multicolumn{2}{c|}{B = 5MB} & \multicolumn{2}{c|}{B = 10MB} \\ \cline{2-9} 
 & \multicolumn{1}{c|}{R@20} & \multicolumn{1}{c|}{N@20} & \multicolumn{1}{c|}{R@20} & N@20 & \multicolumn{1}{c|}{R@20} & \multicolumn{1}{c|}{N@20} & \multicolumn{1}{c|}{R@20} & N@20 \\ \hline
ES & \multicolumn{1}{c|}{0.0474} & \multicolumn{1}{c|}{0.0256} & \multicolumn{1}{c|}{0.0520} & 0.0279 & \multicolumn{1}{c|}{0.0417} & \multicolumn{1}{c|}{0.0231} & \multicolumn{1}{c|}{0.0475} & 0.0263 \\ \hline
MR & \multicolumn{1}{c|}{0.0386} & \multicolumn{1}{c|}{0.0210} & \multicolumn{1}{c|}{0.0496} & 0.0270 & \multicolumn{1}{c|}{0.0312} & \multicolumn{1}{c|}{0.0175} & \multicolumn{1}{c|}{0.0356} & 0.0196 \\ \hline
CIESS & \multicolumn{1}{c|}{0.0502} & \multicolumn{1}{c|}{0.0278} & \multicolumn{1}{c|}{0.0538} & 0.0296 & \multicolumn{1}{c|}{0.0414} & \multicolumn{1}{c|}{0.0233} & \multicolumn{1}{c|}{0.0483} & 0.0274 \\ \hline
BET & \multicolumn{1}{c|}{0.0524} & \multicolumn{1}{c|}{0.0289} & \multicolumn{1}{c|}{0.0557} & 0.0204 & \multicolumn{1}{c|}{0.0439} & \multicolumn{1}{c|}{0.0248} & \multicolumn{1}{c|}{0.0495} & 0.0280 \\ \hline
SCALL & \multicolumn{1}{c|}{\textbf{0.0545}} & \multicolumn{1}{c|}{\textbf{0.0300}} & \multicolumn{1}{c|}{\textbf{0.0579}} & \textbf{0.0316} & \multicolumn{1}{c|}{\textbf{0.0462}} & \multicolumn{1}{c|}{\textbf{0.0259}} & \multicolumn{1}{c|}{\textbf{0.0515}} & \textbf{0.0291} \\ \hline
 & \multicolumn{2}{c|}{B} & \multicolumn{1}{c|}{R@20} & N@20 & \multicolumn{2}{c|}{B} & \multicolumn{1}{c|}{R@20} & N@20 \\ \hline
ESAPN & \multicolumn{2}{c|}{9.29MB} & \multicolumn{1}{c|}{0.0539} & 0.0300 & \multicolumn{2}{c|}{8.12MB} & \multicolumn{1}{c|}{0.0442} & 0.0248 \\ \hline
DESS & \multicolumn{2}{c|}{7.66MB} & \multicolumn{1}{c|}{0.0520} & 0.0287 & \multicolumn{2}{c|}{6.30MB} & \multicolumn{1}{c|}{0.0450} & 0.0252 \\ \hline
AutoEmb & \multicolumn{2}{c|}{7.98MB} & \multicolumn{1}{c|}{0.0505} & 0.0278 & \multicolumn{2}{c|}{9.76MB} & \multicolumn{1}{c|}{0.0458} & 0.0258 \\ \hline
\end{tabular}%
}
\label{tab:fixedb}
\end{table*}

\section{Experiments}
In this section, we elucidate our experimental analysis on SCALL. Through the experiments, we seek to answer the following research questions: 
\begin{itemize}
    \item \textbf{RQ1}: how does SCALL perform compared to ES and other embedding size search methods in streaming settings without the need to be retrained?
    \item \textbf{RQ2}: Can SCALL enforce any predefined memory budget during embedding size search?
    \item \textbf{RQ3}: Can SCALL dynamically adjust embedding sizes for users and items?
    \item \textbf{RQ4}: How does the reservoir size $\gamma$ impact the performance?
    \item \textbf{RQ5}: Which probabilistic distribution yields the best performance?
\end{itemize}

\subsection{Baseline Methods}
To verify the superiority of the proposed method, we compare SCALL against several embedding size search algorithms that can be classified into three groups: fixed embedding size (i.e., ES), static methods (i.e., MR, CIESS and BET) and dynamic methods (i.e., ESAPN, DESS, AutoEmb and SCALL) for streaming recommendation. In static methods, the user and item embedding sizes remain fixed. We retrain them from scratch in every data segment so they are comparable to dynamic methods, which adjusts the embedding sizes in each time segment. The baseline methods are listed below:
\begin{itemize}
    \item CIESS \cite{ciess}: A TD3-based embedding size search algorithm that chooses the embedding sizes from a continuous domain.
    \item BET \cite{bet}: A neural architecture search method that samples numerous table-level candidate embedding sizes and selects the best action using a DeepSets-based fitness predictor.
    \item AutoEmb \cite{autoemb}: A differentiable framework that can automatically adjust embedding sizes based on user/item popularity in a streaming setting.
    \item ESAPN \cite{esapn}: A reinforcement learning-based embedding size search algorithm that performs hard selection on embedding sizes based on memory complexity and recommendation performance.
    \item DESS \cite{dess}: A non-stationary LinUCB-based method with a sublinear regret upper bound and solves the problem of embedding size search by minimizing the cumulative regret.
    \item  Equal Sizes (ES): ES assigns equal embedding sizes for all users and items.
    \item Mixed and Random (MR): MR samples random embedding sizes  from a uniform distribution.
\end{itemize}

\subsection{Datasets and Evaluation Protocols}
The proposed and baseline method undergo evaluation on two real-world datasets: Amazon-Book \cite{amazondataset} and Yelp \footnote{https://business.yelp.com/data/resources/open-dataset/}, both of which contain user ratings. We apply 30-core and 10-core processing on the two datasets, respectively. All the remaining interactions are treated as positive and are sorted in chronological order. After processing, the former dataset comprises 5,076,971 interactions involving 66,351 users and 57,270 items, while the latter involves 2,762,098 interactions between 99,011 users and 56,441 items. A new data segment is created whenever 15,000 users or items are introduced (i.e., $m = 15,000$). We partition each data segment into 80\% for training and 20\% for testing, ensuring that the training data precedes the test data chronologically.  As a result, 10 data segments are created for the Yelp dataset and 8 data segments are created for the Amazon-Book dataset. 

We employ Recall@20 and NDCG@20 as evaluation metrics. The final performance of each embedding sparsification method is evaluated using the mean Recall@20 and NDCG@20 scores over every data segment.

We also test SCALL under two scenarios - fixed mean embedding sizes and fixed numbers of parameters. In the first scenario, we set the mean embedding size $c$ to 16, 32, and 64. In second scenario, the number of parameters $B$ is set to 1,250,000 and 2,500,000, corresponding to 5MB and 10MB memory. It is worth pointing out that memory budgets larger than 10MB (e.g., 15MB) are likely to cause the mean embedding size to exceed 256, which is the full embedding size in our setting, and is thus unnecessary. However, as AutoEmb, ESAPN and DESS are not explicitly designed for precise control over embedding sparsity and prioritize performance, this section solely presents their performance linked to the final embedding tables after pruning.

\subsection{Implementation Details}
In SCALL, we iteratively tune the base recommender to search for the optimal embedding sizes. Once the policy network converges, we train the base recommender till it converges. LightGCN \cite{he2020lightgcn} is employed as the base recommender. During the training and tuning phases, we utilize a batch size of 10,000. The base recommender is updated 200 times each time it is tuned. During the training phase, learning rate decay is applied. The initial learning rate is set to 0.03 and decays every 200 training steps, with 0.95 decay rate. The minimum learning rate set to 0.001. The full embedding size $d_{max}$ is 256 and the minimal embedding size $d_{min}$ is 1. Early stopping is integrated into the training phase, with evaluation occurring every 200 steps of updates and an early stopping patience of 3. 

When training the policy network, $\alpha_{max}$ and $\alpha_{min}$ are set to 30 and 0.1 for Power Law distributions, $10 \times 10^{-5}$ and $10 \times 10^{-6}$ for normal distributions, 100 and 30 for Pareto distributions. The user and item interaction counts are each divided into $L = 256$ groups when we calculate $\boldsymbol{f}^U$ and $\boldsymbol{f}^V$. Both the actor and critic networks are MLPs with one hidden layer and the hidden size is set to 512. 8 transition tuples are sampled from the replay buffer in each iteration.

\subsection{Overall Performance Comparison}

Table \ref{tab:fixedc} and Table \ref{tab:fixedb} present the Recall@20 and NDCG@20 scores obtained from various embedding sparsification methods with fixed mean embedding sizes and fixed numbers of parameters, respectively. Additionally, we conduct paired Student's $t$-tests between SCALL and the best-performing baseline method in each setting. As shown in both tables, across sparsity levels ($c \in \{16, 32, 64\}$ and $B \in \{5\text{MB}, 10\text{MB}\}$), SCALL consistently outperforms the baselines with adjustable sparsity ratios (i.e., ES, MR, BET, and CIESS) in all settings. MR has the worst performance in all settings. Notably, as shown by Table \ref{tab:fixedc}, at a higher sparsification level, SCALL with $c = 64$ surpasses ESAPN, DESS and AutoEmb with higher memory usage on both datasets, which proves the superiority of SCALL. Table \ref{tab:fixedb} also shows that, although using more parameters ($B > 5\text{MB}$), ESAPN, DESS and AutoEmb do not outperform SCALL with $B = 5\text{MB}$. The associated $p$-values, all below 0.05, provide statistical significance, indicating that the advantageous efficacy of SCALL results are not random and thus answering RQ1. Tables \ref{tab:fixedc} and \ref{tab:fixedb} further demonstrate that, unlike ESAPN, DESS and AutoEmb, SCALL can enforce any memory budget, thus addressing RQ2. It's worth noting that, through tuning the scaling coefficient in its reward function, CIESS also achieves memory budget enforcement \cite{ciess}, but this approach is often impractical in real-world implementations.

It is important to highlight that while CIESS and BET demonstrate respectable performance across various settings, their lack of compatibility with streaming recommendation settings entails initialization and training from scratch in each data segment. Consequently, the associated computational costs render them impractical for large-scale recommendation applications.

\subsection{Model Component Analysis}
\subsubsection{User-item Separation} As described in section \ref{subsec:action}, we separate users from items and sample their embedding sizes from different distributions due to their distributional difference. To verify the performance benefits stemming from this design choice, we conduct an experiment where we aggregate users and items into a unified set of $L$ groups, ordered by descending frequency in $D_t^{tr} \cup \mathcal{R}$. We also sample the embedding sizes from one single Power Law distribution parameterized by $\alpha$. The state and action are simplified as follows:
\begin{align}
    s &= (\boldsymbol{f}, r, \frac{|\mathcal{U}|}{|\mathcal{U}|+|\mathcal{V}|}, \boldsymbol{m}^U, \boldsymbol{m}^V) \nonumber \\
    a &= \alpha,
\end{align}
where $\boldsymbol{f}$ contains the sorted normalized frequency for all users and items. The mean embedding size $c$ is 32. We refer to this simplified version of SCALL as Dummy and compare its performance against the standard version, which maintains user item separation. The results, presented in Table \ref{tab:variation}, demonstrate the efficacy of separately grouping users and items in enhancing performance.

\subsubsection{State Information} As outlined in Section \ref{subsec:state}, the state representation incorporates information about the statistical dispersion of resulting embedding sizes, encapsulated in their normalized range denoted as $h^U$ and $h^V$ in each state. Additionally, to capture the latent information within each data segment and determine the corresponding embedding sizes, we calculate the mean user embedding vectors $\boldsymbol{m}^U$ across all users and the mean item embedding vectors $\boldsymbol{m}^V$ across all items. To assess the impact of $h^U$ and $h^V$ on the final recommendation quality, we exclude them from the state representation and report the Recall@20 and NDCG@20 scores. Similarly, we omit $\boldsymbol{m}^U$ and $\boldsymbol{m}^V$ from the state representation before algorithm execution and report the same metrics. We set the mean embedding size $c$ to 32. The results are displayed in Table \ref{tab:state}. By comparing these results with the performance of the standard version of SCALL, we conclude that removing any of these components results in a degradation of performance.

\begin{table}[]
\caption{Performance of different variations of SCALL after one of the state components is removed.}
\resizebox{0.8\columnwidth}{!}{%
\begin{tabular}{|c|cc|cc|}
\hline
 & \multicolumn{2}{c|}{Yelp} & \multicolumn{2}{c|}{Amazon-Book} \\ \hline
\begin{tabular}[c]{@{}c@{}}Variation\end{tabular} & \multicolumn{1}{c|}{R@20} & N@20 & \multicolumn{1}{c|}{R@20} & N@20 \\ \hline
w/o $\boldsymbol{m}^{U/V}$ & \multicolumn{1}{c|}{0.0594} & 0.0324 & \multicolumn{1}{c|}{0.0510} & 0.0286 \\ \hline
w/o $h^{U/V}$ & \multicolumn{1}{c|}{0.0589} & 0.0323 & \multicolumn{1}{c|}{0.0507} & 0.0284 \\ \hline
Dummy & \multicolumn{1}{c|}{0.0501} & 0.0268 & \multicolumn{1}{c|}{0.0387} & 0.0217 \\ \hline
\end{tabular}%
}

\label{tab:state}
\end{table}

\begin{table}[] 
\caption{Performance metrics of SCALL variations adopting different distributions.}
\resizebox{0.8\columnwidth}{!}{%
\begin{tabular}{|c|cc|cc|}
\hline
 & \multicolumn{2}{c|}{Yelp} & \multicolumn{2}{c|}{Amazon-Book} \\ \hline
Distribution & \multicolumn{1}{c|}{R@20} & N@20 & \multicolumn{1}{c|}{R@20} & N@20 \\ \hline
Pareto & \multicolumn{1}{c|}{0.0602} & 0.0325 & \multicolumn{1}{c|}{0.0512} & 0.0287 \\ \hline
Normal & \multicolumn{1}{c|}{0.0602} & 0.0329 & \multicolumn{1}{c|}{0.0504} & 0.0285 \\ \hline
Power Law & \multicolumn{1}{c|}{0.0604} & 0.0329 & \multicolumn{1}{c|}{0.0516} & 0.0290 \\ \hline
\end{tabular}%
}
\label{tab:variation}
\end{table}

\subsection{Analysis of Hyperparameters}
In this section, we examine the impact of fundamental hyperparameters within SCALL w.r.t. their influence on Recall@20 and NDCG@20.

\subsubsection{Reservoir Size $\gamma$}
\begin{figure}
    \centering
    \includegraphics[width=\linewidth]{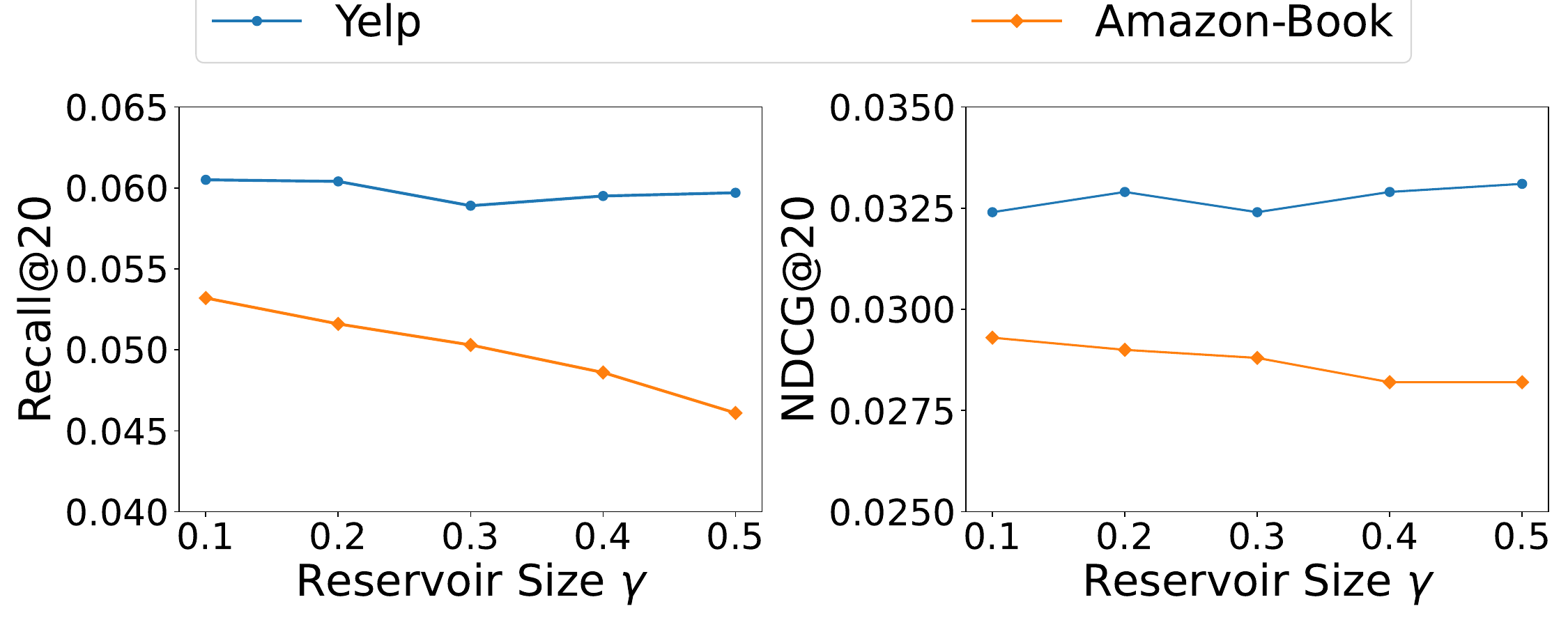} 
    \caption{Sensitivity analysis of $\gamma$ w.r.t. Recall@20 and NDCG@20 on the Yelp and Amazon-Book dataset.} \label{fig:gamma}
\end{figure}

During time segment $t$, the base recommender undergoes training on the combined dataset $D^{tr}_t \cup \mathcal{R}_t$, where $\mathcal{R}_t$ represents a reservoir containing historical user-item interactions sampled from $D_1 \cup ... \cup D_{t-2} \cup D_{t-1}$. The reservoir's size is determined as $\gamma |D^{tr}_t|$, where $|D^{tr}_t|$ represents the number of interactions in $D^{tr}_t$. Users and items are sorted based on their frequency in $D^{tr}_t \cup \mathcal{R}_t$, allowing for the adjustment of the influence that past records exert on the current embedding size search. To answer RQ4, we explore the impact of different $\gamma$ values from $\{0.1, 0.2, 0.3, 0.4, 0.5\}$ on Recall@20 and NDCG@20 scores to investigate how the reservoir size affects recommendation performance. The mean embedding size $c$ is set to 32. The results, depicted in Figure \ref{fig:gamma}, reveal that SCALL achieves optimal performance when $\gamma$ is set to $0.1$ for Amazon-Book. The performance of SCALL is generally insensitive to the setting of $\gamma$ on the Yelp dataset, showcasing no significant difference with varying values of $\gamma$. 

\subsubsection{Choice of Distributions}
As outlined in Section \ref{subsec:action}, the distribution of parameters among user and item groups is determined by sampling values from two Power Law distributions. To answer RQ5, we replace the Power Law distributions with truncated normal distributions and Pareto distributions, which are commonly used to model social income and wealth distribution. Similar to the Power Law distributions, both types of distributions are parameterized by $\alpha_U$ and $\alpha_V$, which is bounded by $(\alpha_{min}, \alpha_{max})$. For the normal distributions, they are bounded within $(0, 1)$. The normal distribution for the users has mean of $\frac{1}{|\mathcal{U}|}$ and the one for the items has mean of $\frac{1}{|\mathcal{V}|}$. Their scales are set to $\alpha^U$ and $\alpha^V$. The Pareto distributions are parameterized by $\alpha^U$ and $\alpha^V$. The fairness of parameter allocation can be adjusted by adjusting the values of $\alpha^U$ and $\alpha^V$. The parameters tend be to distributed evenly or to the most frequent users and items when $\alpha^U$ and $\alpha^V$ approach one of the two extremes. The mean embedding size $c$ is set to 32. The performance comparison presented in Table \ref{tab:variation} indicates that SCALL is generally insensitive to the choice of distributions, with Power Law and Pareto distributions achieving slightly better performance.

\subsection{Case Study of User Embedding Sizes}
\begin{figure}
    \centering
    \includegraphics[width=\linewidth]{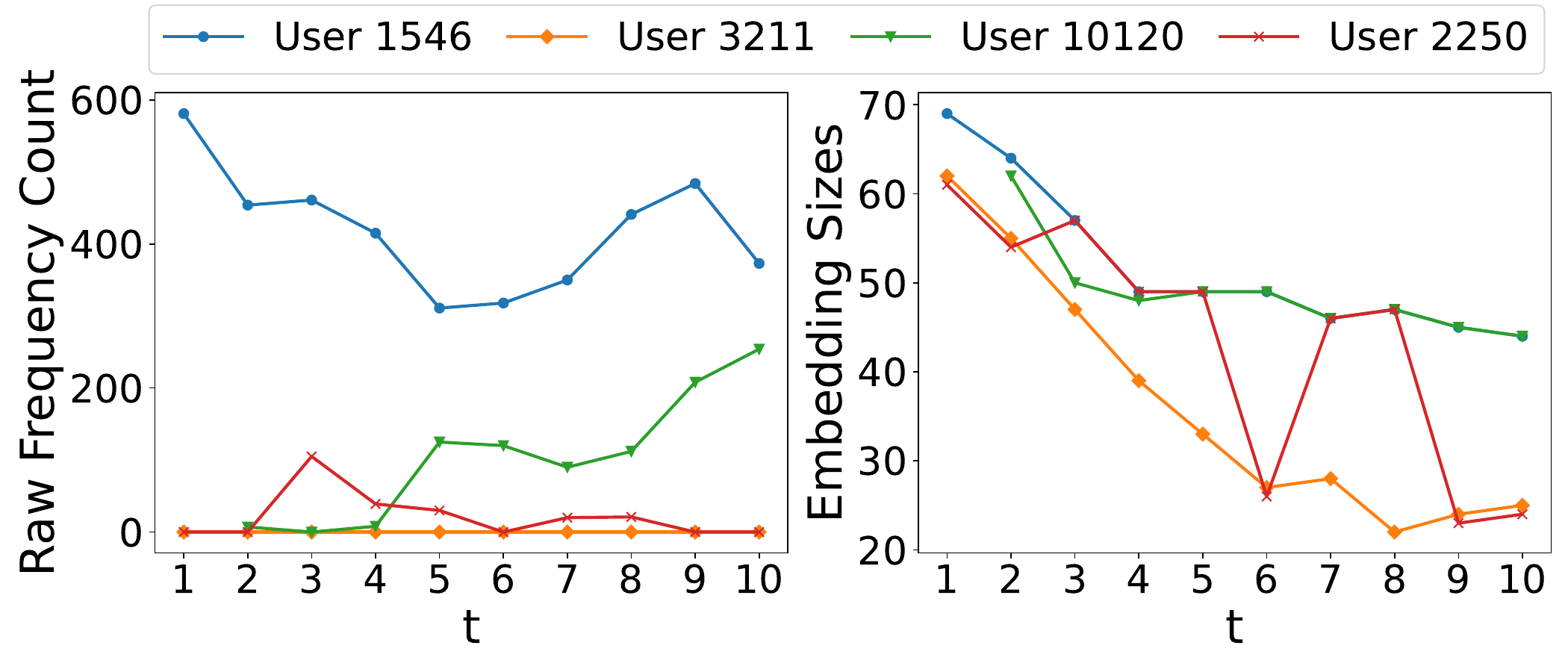} 
    \caption{Case study on the interaction counts (left) and embedding sizes (right) of four users from the Yelp dataset.} \label{fig:case}
    \vspace{-1em}
\end{figure}

To gain insight into how the embedding size varies for specific users or items across different time segments, we focus on four users: user 1546, user 2250, user 3211 and user 10120. We execute the algorithm on the Yelp dataset with the mean embedding size set to 32. Then we generate plots illustrating their interaction counts, which represents their frequency, and corresponding embedding sizes. Each of these users represents a distinct user profile. Users are represented in the plots only after their initial appearance in the dataset. User 1546 consistently maintains high frequency across all segments. In contrast, user 3211 remains unpopular throughout all segments. User 10120 initially starts with low frequency in the first four segments but experiences a steady increase in frequency over the subsequent six segments. Conversely, user 2250 experiences a rise in frequency from the second to the first time segment, followed decline in the subsequent three segments and a slight rise over the seventh and the eighth segments.

As depicted in Figure \ref{fig:case}, all four users initially start with large embedding sizes despite their significant differences in frequency. This phenomenon arises because the number of users is relatively small in the initial segments, allowing each user to receive a relatively generous allocation of embedding sizes. However, user 2250 and user 3211 are still given the smallest embedding size, reflecting its lower frequency compared to the others.

Over time, the embedding size of user 1546 remains consistently large due to its sustained frequency. In contrast, the embedding sizes of users 3211 and 10120 begin to lag behind user 1546 in the first three segments, reflecting their lower frequency. However, while user 10120 initially follows a similar trend as user 3211, its embedding size quickly catches up after the third segment due to an increase in frequency. The embedding size of user 2250 increases in the second segment and decreases in the following three segments, which is consistent with the trend of its frequency.  This frequency later experiences a slight rise in the seventh and eighth segments, leading to an increase in its embedding size. Comparing the embedding sizes of user 10120 and user 1546 after the fourth segment, we find they share the same embedding size although user 1546 is much more frequent. This reveals that SCALL effectively prevents any individual user from receiving an excessive number of parameters by capping the embedding size of highly frequent users. This conclusion is reinforced by examining the frequency and embedding size patterns of users 1546 and 2250 in the seventh and eighth segments.

Furthermore, despite user 3211 having equal frequency to user 2250 in the last two segments, its embedding size is slightly larger. This discrepancy arises because SCALL prioritizes newer users or items when they have the same frequency. Consequently, user 3211, being a newer addition to the dataset compared to user 2250, receives a larger embedding size allotment.

In conclusion, SCALL is able to dynamically adjust the embedding sizes of users and items according to their changing frequency across different data segments, which answers RQ3. 

\section{Conclusion and Future Research}
Recommender Systems face a significant challenge in the form of the memory bottleneck resulting from uniform embedding sizes for all users and items. Existing dynamic embedding size strategies aim to minimize memory footprint but do not have any fixed constraint on the parameter budget. To address these issues, we proposed SCALL, an automated policy that dynamically adjusts embedding sizes for each user and item and ensures adherence to a predefined budget. The experimental results on the Yelp and Amazon-Book datasets verified the advantageous efficacy of SCALL. However, performing embedding size assignment based on the user and item frequency is biased against the users and items with low frequency. Moving forward, we will concentrate on fair embedding size allocation that overcomes such frequency bias using other importance signals such as model confidence \cite{qu2021human, qu2022combining}.

\section*{Acknowledgment} This work is supported by Australian Research Council under the streams of Future Fellowship (Grant No. FT210100624), Linkage Project (Grant No. LP230200892), Discovery Early Career Researcher Award (Grants No. DE230101033), Discovery Project (Grants No. DP240101108 and No. DP240101814).

\normalem
\bibliographystyle{ACM-Reference-Format}
\bibliography{SCALL/scall}


\begin{thebibliography}{56}


\ifx \showCODEN    \undefined \def \showCODEN     #1{\unskip}     \fi
\ifx \showDOI      \undefined \def \showDOI       #1{#1}\fi
\ifx \showISBNx    \undefined \def \showISBNx     #1{\unskip}     \fi
\ifx \showISBNxiii \undefined \def \showISBNxiii  #1{\unskip}     \fi
\ifx \showISSN     \undefined \def \showISSN      #1{\unskip}     \fi
\ifx \showLCCN     \undefined \def \showLCCN      #1{\unskip}     \fi
\ifx \shownote     \undefined \def \shownote      #1{#1}          \fi
\ifx \showarticletitle \undefined \def \showarticletitle #1{#1}   \fi
\ifx \showURL      \undefined \def \showURL       {\relax}        \fi
\providecommand\bibfield[2]{#2}
\providecommand\bibinfo[2]{#2}
\providecommand\natexlab[1]{#1}
\providecommand\showeprint[2][]{arXiv:#2}

\bibitem[Chang et~al\mbox{.}(2017)]%
        {chang2017www}
\bibfield{author}{\bibinfo{person}{Shiyu Chang}, \bibinfo{person}{Yang Zhang}, \bibinfo{person}{Jiliang Tang}, \bibinfo{person}{Dawei Yin}, \bibinfo{person}{Yi Chang}, \bibinfo{person}{Mark~A. Hasegawa-Johnson}, {and} \bibinfo{person}{Thomas~S. Huang}.} \bibinfo{year}{2017}\natexlab{}.
\newblock \showarticletitle{Streaming Recommender Systems}. In \bibinfo{booktitle}{\emph{WWW}}. \bibinfo{pages}{381–389}.
\newblock


\bibitem[Chen et~al\mbox{.}(2013)]%
        {terec}
\bibfield{author}{\bibinfo{person}{Chen Chen}, \bibinfo{person}{Hongzhi Yin}, \bibinfo{person}{Junjie Yao}, {and} \bibinfo{person}{Bin Cui}.} \bibinfo{year}{2013}\natexlab{}.
\newblock \showarticletitle{TeRec: a temporal recommender system over tweet stream}.
\newblock \bibinfo{journal}{\emph{VLDB}} \bibinfo{volume}{6}, \bibinfo{number}{12} (\bibinfo{year}{2013}), \bibinfo{pages}{1254–1257}.
\newblock


\bibitem[Chen et~al\mbox{.}(2021)]%
        {chentongkdd2021}
\bibfield{author}{\bibinfo{person}{Tong Chen}, \bibinfo{person}{Hongzhi Yin}, \bibinfo{person}{Yujia Zheng}, \bibinfo{person}{Zi Huang}, \bibinfo{person}{Yang Wang}, {and} \bibinfo{person}{Meng Wang}.} \bibinfo{year}{2021}\natexlab{}.
\newblock \showarticletitle{Learning Elastic Embeddings for Customizing On-Device Recommenders}. In \bibinfo{booktitle}{\emph{KDD}}. \bibinfo{pages}{138–147}.
\newblock


\bibitem[Cheng et~al\mbox{.}(2016)]%
        {cheng2016wide}
\bibfield{author}{\bibinfo{person}{Heng-Tze Cheng}, \bibinfo{person}{Levent Koc}, \bibinfo{person}{Jeremiah Harmsen}, \bibinfo{person}{Tal Shaked}, \bibinfo{person}{Tushar Chandra}, \bibinfo{person}{Hrishi Aradhye}, \bibinfo{person}{Glen Anderson}, \bibinfo{person}{Greg Corrado}, \bibinfo{person}{Wei Chai}, \bibinfo{person}{Mustafa Ispir}, {et~al\mbox{.}}} \bibinfo{year}{2016}\natexlab{}.
\newblock \showarticletitle{Wide \& deep learning for recommender systems}. In \bibinfo{booktitle}{\emph{Proceedings of the 1st workshop on deep learning for recommender systems}}. \bibinfo{pages}{7--10}.
\newblock


\bibitem[Guo et~al\mbox{.}(2019)]%
        {sssreckdd2019}
\bibfield{author}{\bibinfo{person}{Lei Guo}, \bibinfo{person}{Hongzhi Yin}, \bibinfo{person}{Qinyong Wang}, \bibinfo{person}{Tong Chen}, \bibinfo{person}{Alexander Zhou}, {and} \bibinfo{person}{Nguyen Quoc Viet~Hung}.} \bibinfo{year}{2019}\natexlab{}.
\newblock \showarticletitle{Streaming Session-Based Recommendation}. In \bibinfo{booktitle}{\emph{KDD}}. \bibinfo{pages}{1569–1577}.
\newblock


\bibitem[Haarnoja et~al\mbox{.}(2018)]%
        {sac}
\bibfield{author}{\bibinfo{person}{Tuomas Haarnoja}, \bibinfo{person}{Aurick Zhou}, \bibinfo{person}{Pieter Abbeel}, {and} \bibinfo{person}{Sergey Levine}.} \bibinfo{year}{2018}\natexlab{}.
\newblock \showarticletitle{Soft actor-critic: Off-policy maximum entropy deep reinforcement learning with a stochastic actor}. In \bibinfo{booktitle}{\emph{ICML}}. \bibinfo{pages}{1861--1870}.
\newblock


\bibitem[He et~al\mbox{.}(2023)]%
        {dess}
\bibfield{author}{\bibinfo{person}{Bowei He}, \bibinfo{person}{Xu He}, \bibinfo{person}{Renrui Zhang}, \bibinfo{person}{Yingxue Zhang}, \bibinfo{person}{Ruiming Tang}, {and} \bibinfo{person}{Chen Ma}.} \bibinfo{year}{2023}\natexlab{}.
\newblock \showarticletitle{Dynamic Embedding Size Search with Minimum Regret for Streaming Recommender System}. In \bibinfo{booktitle}{\emph{CIKM}}. \bibinfo{pages}{741–750}.
\newblock


\bibitem[He et~al\mbox{.}(2020)]%
        {he2020lightgcn}
\bibfield{author}{\bibinfo{person}{Xiangnan He}, \bibinfo{person}{Kuan Deng}, \bibinfo{person}{Xiang Wang}, \bibinfo{person}{Yan Li}, \bibinfo{person}{Yongdong Zhang}, {and} \bibinfo{person}{Meng Wang}.} \bibinfo{year}{2020}\natexlab{}.
\newblock \showarticletitle{Lightgcn: Simplifying and powering graph convolution network for recommendation}. In \bibinfo{booktitle}{\emph{SIGIR}}. \bibinfo{pages}{639--648}.
\newblock


\bibitem[He et~al\mbox{.}(2017)]%
        {he2017neural}
\bibfield{author}{\bibinfo{person}{Xiangnan He}, \bibinfo{person}{Lizi Liao}, \bibinfo{person}{Hanwang Zhang}, \bibinfo{person}{Liqiang Nie}, \bibinfo{person}{Xia Hu}, {and} \bibinfo{person}{Tat-Seng Chua}.} \bibinfo{year}{2017}\natexlab{}.
\newblock \showarticletitle{Neural collaborative filtering}. In \bibinfo{booktitle}{\emph{WWW}}. \bibinfo{pages}{173--182}.
\newblock


\bibitem[Joglekar et~al\mbox{.}(2020)]%
        {joglekar2020neural}
\bibfield{author}{\bibinfo{person}{Manas~R Joglekar}, \bibinfo{person}{Cong Li}, \bibinfo{person}{Mei Chen}, \bibinfo{person}{Taibai Xu}, \bibinfo{person}{Xiaoming Wang}, \bibinfo{person}{Jay~K Adams}, \bibinfo{person}{Pranav Khaitan}, \bibinfo{person}{Jiahui Liu}, {and} \bibinfo{person}{Quoc~V Le}.} \bibinfo{year}{2020}\natexlab{}.
\newblock \showarticletitle{Neural input search for large scale recommendation models}. In \bibinfo{booktitle}{\emph{KDD}}. \bibinfo{pages}{2387--2397}.
\newblock


\bibitem[Kong et~al\mbox{.}(2023)]%
        {autosrh}
\bibfield{author}{\bibinfo{person}{Shuming Kong}, \bibinfo{person}{Weiyu Cheng}, \bibinfo{person}{Yanyan Shen}, {and} \bibinfo{person}{Linpeng Huang}.} \bibinfo{year}{2023}\natexlab{}.
\newblock \showarticletitle{AutoSrh: An Embedding Dimensionality Search Framework for Tabular Data Prediction}.
\newblock \bibinfo{journal}{\emph{TKDE}} \bibinfo{volume}{35}, \bibinfo{number}{7} (\bibinfo{year}{2023}), \bibinfo{pages}{6673--6686}.
\newblock


\bibitem[Kywe et~al\mbox{.}(2012)]%
        {kywe2012survey}
\bibfield{author}{\bibinfo{person}{Su~Mon Kywe}, \bibinfo{person}{Ee-Peng Lim}, {and} \bibinfo{person}{Feida Zhu}.} \bibinfo{year}{2012}\natexlab{}.
\newblock \showarticletitle{A survey of recommender systems in twitter}. In \bibinfo{booktitle}{\emph{International Conference on Social Informatics}}. \bibinfo{pages}{420--433}.
\newblock


\bibitem[Lee et~al\mbox{.}(2018)]%
        {lee2018collaborative}
\bibfield{author}{\bibinfo{person}{Joonseok Lee}, \bibinfo{person}{Sami Abu-El-Haija}, \bibinfo{person}{Balakrishnan Varadarajan}, {and} \bibinfo{person}{Apostol Natsev}.} \bibinfo{year}{2018}\natexlab{}.
\newblock \showarticletitle{Collaborative deep metric learning for video understanding}. In \bibinfo{booktitle}{\emph{KDD}}. \bibinfo{pages}{481--490}.
\newblock


\bibitem[Li et~al\mbox{.}(2023)]%
        {jiachengli2023tia}
\bibfield{author}{\bibinfo{person}{Jiacheng Li}, \bibinfo{person}{Ming Wang}, \bibinfo{person}{Jin Li}, \bibinfo{person}{Jinmiao Fu}, \bibinfo{person}{Xin Shen}, \bibinfo{person}{Jingbo Shang}, {and} \bibinfo{person}{Julian McAuley}.} \bibinfo{year}{2023}\natexlab{}.
\newblock \showarticletitle{Text Is All You Need: Learning Language Representations for Sequential Recommendation}. In \bibinfo{booktitle}{\emph{KDD}}. \bibinfo{pages}{1258–1267}.
\newblock


\bibitem[Li and Karahanna(2015)]%
        {rececom1}
\bibfield{author}{\bibinfo{person}{Siyuan Li} {and} \bibinfo{person}{Elena Karahanna}.} \bibinfo{year}{2015}\natexlab{}.
\newblock \showarticletitle{Online Recommendation Systems in a B2C E-Commerce Context: A Review and Future Directions}.
\newblock \bibinfo{journal}{\emph{Journal of the Association for Information Systems}} \bibinfo{volume}{16}, \bibinfo{number}{2} (\bibinfo{year}{2015}), \bibinfo{pages}{72--107}.
\newblock


\bibitem[Liang et~al\mbox{.}(2024)]%
        {liang2024legcf}
\bibfield{author}{\bibinfo{person}{Xurong Liang}, \bibinfo{person}{Tong Chen}, \bibinfo{person}{Lizhen Cui}, \bibinfo{person}{Yang Wang}, \bibinfo{person}{Meng Wang}, {and} \bibinfo{person}{Hongzhi Yin}.} \bibinfo{year}{2024}\natexlab{}.
\newblock \showarticletitle{Lightweight Embeddings for Graph Collaborative Filtering}. In \bibinfo{booktitle}{\emph{SIGIR}}. \bibinfo{pages}{1296–1306}.
\newblock


\bibitem[Liang et~al\mbox{.}(2023)]%
        {liang2023lcc}
\bibfield{author}{\bibinfo{person}{Xurong Liang}, \bibinfo{person}{Tong Chen}, \bibinfo{person}{Quoc Viet~Hung Nguyen}, \bibinfo{person}{Jianxin Li}, {and} \bibinfo{person}{Hongzhi Yin}.} \bibinfo{year}{2023}\natexlab{}.
\newblock \showarticletitle{Learning Compact Compositional Embeddings via Regularized Pruning for Recommendation}. In \bibinfo{booktitle}{\emph{ICDM}}. \bibinfo{pages}{378--387}.
\newblock


\bibitem[Lin et~al\mbox{.}(2022)]%
        {adafs}
\bibfield{author}{\bibinfo{person}{Weilin Lin}, \bibinfo{person}{Xiangyu Zhao}, \bibinfo{person}{Yejing Wang}, {and} \bibinfo{person}{Xian Xu, Tong ad~Wu}.} \bibinfo{year}{2022}\natexlab{}.
\newblock \showarticletitle{AdaFS: Adaptive Feature Selection in Deep Recommender System}. In \bibinfo{booktitle}{\emph{KDD}}. \bibinfo{pages}{3309–3317}.
\newblock


\bibitem[Liu et~al\mbox{.}(2020)]%
        {esapn}
\bibfield{author}{\bibinfo{person}{Haochen Liu}, \bibinfo{person}{Xiangyu Zhao}, \bibinfo{person}{Chong Wang}, \bibinfo{person}{Xiaobing Liu}, {and} \bibinfo{person}{Jiliang Tang}.} \bibinfo{year}{2020}\natexlab{}.
\newblock \showarticletitle{Automated embedding size search in deep recommender systems}. In \bibinfo{booktitle}{\emph{SIGIR}}. \bibinfo{pages}{2307--2316}.
\newblock


\bibitem[Liu et~al\mbox{.}(2021)]%
        {liu2021learnable}
\bibfield{author}{\bibinfo{person}{Siyi Liu}, \bibinfo{person}{Chen Gao}, \bibinfo{person}{Yihong Chen}, \bibinfo{person}{Depeng Jin}, {and} \bibinfo{person}{Yong Li}.} \bibinfo{year}{2021}\natexlab{}.
\newblock \showarticletitle{Learnable Embedding sizes for Recommender Systems}. In \bibinfo{booktitle}{\emph{ICLR}}.
\newblock


\bibitem[Luo et~al\mbox{.}(2019)]%
        {luo2019autocross}
\bibfield{author}{\bibinfo{person}{Yuanfei Luo}, \bibinfo{person}{Mengshuo Wang}, \bibinfo{person}{Hao Zhou}, \bibinfo{person}{Quanming Yao}, \bibinfo{person}{Wei-Wei Tu}, \bibinfo{person}{Yuqiang Chen}, \bibinfo{person}{Wenyuan Dai}, {and} \bibinfo{person}{Qiang Yang}.} \bibinfo{year}{2019}\natexlab{}.
\newblock \showarticletitle{Autocross: Automatic feature crossing for tabular data in real-world applications}. In \bibinfo{booktitle}{\emph{KDD}}. \bibinfo{pages}{1936--1945}.
\newblock


\bibitem[Lyu et~al\mbox{.}(2022)]%
        {optembed}
\bibfield{author}{\bibinfo{person}{Fuyuan Lyu}, \bibinfo{person}{Xing Tang}, \bibinfo{person}{Hong Zhu}, \bibinfo{person}{Huifeng Guo}, \bibinfo{person}{Yingxue Zhang}, \bibinfo{person}{Ruiming Tang}, {and} \bibinfo{person}{Xue Liu}.} \bibinfo{year}{2022}\natexlab{}.
\newblock \showarticletitle{OptEmbed: Learning Optimal Embedding Table for Click-through Rate Prediction}. In \bibinfo{booktitle}{\emph{CIKM}}. \bibinfo{pages}{1399–1409}.
\newblock


\bibitem[Mao et~al\mbox{.}(2021)]%
        {ultragcn}
\bibfield{author}{\bibinfo{person}{Kelong Mao}, \bibinfo{person}{Jieming Zhu}, \bibinfo{person}{Xi Xiao}, \bibinfo{person}{Biao Lu}, \bibinfo{person}{Zhaowei Wang}, {and} \bibinfo{person}{Xiuqiang He}.} \bibinfo{year}{2021}\natexlab{}.
\newblock \showarticletitle{UltraGCN: Ultra Simplification of Graph Convolutional Networks for Recommendation}. In \bibinfo{booktitle}{\emph{CIKM}}. \bibinfo{pages}{1253–1262}.
\newblock


\bibitem[Ni et~al\mbox{.}(2019)]%
        {amazondataset}
\bibfield{author}{\bibinfo{person}{Jianmo Ni}, \bibinfo{person}{Jiacheng Li}, {and} \bibinfo{person}{Julian McAuley}.} \bibinfo{year}{2019}\natexlab{}.
\newblock \showarticletitle{Justifying Recommendations using Distantly-Labeled Reviews and Fine-Grained Aspects}. In \bibinfo{booktitle}{\emph{Proceedings of the 2019 Conference on Empirical Methods in Natural Language Processing and the 9th International Joint Conference on Natural Language Processing (EMNLP-IJCNLP)}}. \bibinfo{pages}{188--197}.
\newblock


\bibitem[Qiu et~al\mbox{.}(2020)]%
        {gag}
\bibfield{author}{\bibinfo{person}{Ruihong Qiu}, \bibinfo{person}{Hongzhi Yin}, \bibinfo{person}{Zi Huang}, {and} \bibinfo{person}{Tong Chen}.} \bibinfo{year}{2020}\natexlab{}.
\newblock \showarticletitle{GAG: Global Attributed Graph Neural Network for Streaming Session-Based Recommendation}. In \bibinfo{booktitle}{\emph{SIGIR}}. \bibinfo{pages}{669–678}.
\newblock


\bibitem[Qu et~al\mbox{.}(2022b)]%
        {singleshot}
\bibfield{author}{\bibinfo{person}{Liang Qu}, \bibinfo{person}{Yonghong Ye}, \bibinfo{person}{Ningzhi Tang}, \bibinfo{person}{Lixin Zhang}, \bibinfo{person}{Yuhui Shi}, {and} \bibinfo{person}{Hongzhi Yin}.} \bibinfo{year}{2022}\natexlab{b}.
\newblock \showarticletitle{Single-Shot Embedding Dimension Search in Recommender System}. In \bibinfo{booktitle}{\emph{SIGIR}}. \bibinfo{pages}{513–522}.
\newblock


\bibitem[Qu et~al\mbox{.}(2024)]%
        {bet}
\bibfield{author}{\bibinfo{person}{Yunke Qu}, \bibinfo{person}{Tong Chen}, \bibinfo{person}{Quoc Viet~Hung Nguyen}, {and} \bibinfo{person}{Hongzhi Yin}.} \bibinfo{year}{2024}\natexlab{}.
\newblock \showarticletitle{Budgeted Embedding Table For Recommender Systems}. In \bibinfo{booktitle}{\emph{WSDM}}. \bibinfo{pages}{557–566}.
\newblock


\bibitem[Qu et~al\mbox{.}(2023)]%
        {ciess}
\bibfield{author}{\bibinfo{person}{Yunke Qu}, \bibinfo{person}{Tong Chen}, \bibinfo{person}{Xiangyu Zhao}, \bibinfo{person}{Lizhen Cui}, \bibinfo{person}{Kai Zheng}, {and} \bibinfo{person}{Hongzhi Yin}.} \bibinfo{year}{2023}\natexlab{}.
\newblock \showarticletitle{Continuous Input Embedding Size Search For Recommender Systems}. In \bibinfo{booktitle}{\emph{SIGIR}}. \bibinfo{pages}{708–717}.
\newblock


\bibitem[Qu et~al\mbox{.}(2022a)]%
        {qu2022combining}
\bibfield{author}{\bibinfo{person}{Yunke Qu}, \bibinfo{person}{Kevin Roitero}, \bibinfo{person}{David~La Barbera}, \bibinfo{person}{Damiano Spina}, \bibinfo{person}{Stefano Mizzaro}, {and} \bibinfo{person}{Gianluca Demartini}.} \bibinfo{year}{2022}\natexlab{a}.
\newblock \showarticletitle{Combining Human and Machine Confidence in Truthfulness Assessment}.
\newblock \bibinfo{journal}{\emph{J. Data and Information Quality}} \bibinfo{volume}{15}, \bibinfo{number}{1} (\bibinfo{year}{2022}).
\newblock


\bibitem[Qu et~al\mbox{.}(2021)]%
        {qu2021human}
\bibfield{author}{\bibinfo{person}{Yunke Qu}, \bibinfo{person}{Kevin Roitero}, \bibinfo{person}{Stefano Mizzaro}, \bibinfo{person}{Damiano Spina}, {and} \bibinfo{person}{Gianluca Demartini}.} \bibinfo{year}{2021}\natexlab{}.
\newblock \showarticletitle{Human-in-the-Loop Systems for Truthfulness: A Study of Human and Machine Confidence}. In \bibinfo{booktitle}{\emph{Conference for Truth and Trust Online}}.
\newblock


\bibitem[Rendle et~al\mbox{.}(2009)]%
        {rendle_bpr_2012}
\bibfield{author}{\bibinfo{person}{Steffen Rendle}, \bibinfo{person}{Christoph Freudenthaler}, \bibinfo{person}{Zeno Gantner}, {and} \bibinfo{person}{Lars Schmidt-Thieme}.} \bibinfo{year}{2009}\natexlab{}.
\newblock \showarticletitle{BPR: Bayesian Personalized Ranking from Implicit Feedback}. In \bibinfo{booktitle}{\emph{Proceedings of the Twenty-Fifth Conference on Uncertainty in Artificial Intelligence}}. \bibinfo{pages}{452–461}.
\newblock


\bibitem[Sedaghati et~al\mbox{.}(2015)]%
        {sedaghati2015automatic}
\bibfield{author}{\bibinfo{person}{Naser Sedaghati}, \bibinfo{person}{Te Mu}, \bibinfo{person}{Louis-Noel Pouchet}, \bibinfo{person}{Srinivasan Parthasarathy}, {and} \bibinfo{person}{P Sadayappan}.} \bibinfo{year}{2015}\natexlab{}.
\newblock \showarticletitle{Automatic selection of sparse matrix representation on GPUs}. In \bibinfo{booktitle}{\emph{ACM International Conference on Supercomputing}}. \bibinfo{pages}{99--108}.
\newblock


\bibitem[Tran et~al\mbox{.}(2024)]%
        {tran2024thoroug}
\bibfield{author}{\bibinfo{person}{Hung~Vinh Tran}, \bibinfo{person}{Tong Chen}, \bibinfo{person}{Quoc Viet~Hung Nguyen}, \bibinfo{person}{Zi Huang}, \bibinfo{person}{Lizhen Cui}, {and} \bibinfo{person}{Hongzhi Yin}.} \bibinfo{year}{2024}\natexlab{}.
\newblock \bibinfo{title}{A Thorough Performance Benchmarking on Lightweight Embedding-based Recommender Systems}.
\newblock
\newblock
\showeprint[arxiv]{2406.17335}


\bibitem[Virtanen et~al\mbox{.}(2020)]%
        {virtanen2020scipy}
\bibfield{author}{\bibinfo{person}{Pauli Virtanen} {et~al\mbox{.}}} \bibinfo{year}{2020}\natexlab{}.
\newblock \showarticletitle{SciPy 1.0: fundamental algorithms for scientific computing in Python}.
\newblock \bibinfo{journal}{\emph{Nature Methods}} \bibinfo{volume}{17}, \bibinfo{number}{3} (\bibinfo{year}{2020}), \bibinfo{pages}{261--272}.
\newblock


\bibitem[Wang et~al\mbox{.}(2017a)]%
        {wang2017als}
\bibfield{author}{\bibinfo{person}{Hao Wang}, \bibinfo{person}{Yanmei Fu}, \bibinfo{person}{Qinyong Wang}, \bibinfo{person}{Hongzhi Yin}, \bibinfo{person}{Changying Du}, {and} \bibinfo{person}{Hui Xiong}.} \bibinfo{year}{2017}\natexlab{a}.
\newblock \showarticletitle{A Location-Sentiment-Aware Recommender System for Both Home-Town and Out-of-Town Users}. In \bibinfo{booktitle}{\emph{KDD}}. \bibinfo{pages}{1135–1143}.
\newblock


\bibitem[Wang et~al\mbox{.}(2020)]%
        {wang2020npo}
\bibfield{author}{\bibinfo{person}{Qinyong Wang}, \bibinfo{person}{Hongzhi Yin}, \bibinfo{person}{Tong Chen}, \bibinfo{person}{Zi Huang}, \bibinfo{person}{Hao Wang}, \bibinfo{person}{Yanchang Zhao}, {and} \bibinfo{person}{Nguyen~Quoc Viet~Hung}.} \bibinfo{year}{2020}\natexlab{}.
\newblock \showarticletitle{Next Point-of-Interest Recommendation on Resource-Constrained Mobile Devices}. In \bibinfo{booktitle}{\emph{WWW}}. \bibinfo{pages}{906–916}.
\newblock


\bibitem[Wang et~al\mbox{.}(2018a)]%
        {qinyong2018kdd}
\bibfield{author}{\bibinfo{person}{Qinyong Wang}, \bibinfo{person}{Hongzhi Yin}, \bibinfo{person}{Zhiting Hu}, \bibinfo{person}{Defu Lian}, \bibinfo{person}{Hao Wang}, {and} \bibinfo{person}{Zi Huang}.} \bibinfo{year}{2018}\natexlab{a}.
\newblock \showarticletitle{Neural Memory Streaming Recommender Networks with Adversarial Training}. In \bibinfo{booktitle}{\emph{KDD}}. \bibinfo{pages}{2467–2475}.
\newblock


\bibitem[Wang et~al\mbox{.}(2021)]%
        {wang2021survey}
\bibfield{author}{\bibinfo{person}{Shoujin Wang}, \bibinfo{person}{Longbing Cao}, \bibinfo{person}{Yan Wang}, \bibinfo{person}{Quan~Z Sheng}, \bibinfo{person}{Mehmet~A Orgun}, {and} \bibinfo{person}{Defu Lian}.} \bibinfo{year}{2021}\natexlab{}.
\newblock \showarticletitle{A survey on session-based recommender systems}.
\newblock \bibinfo{journal}{\emph{ACM Computing Surveys (CSUR)}} \bibinfo{volume}{54}, \bibinfo{number}{7} (\bibinfo{year}{2021}), \bibinfo{pages}{1--38}.
\newblock


\bibitem[Wang et~al\mbox{.}(2017b)]%
        {suhangwang2017www}
\bibfield{author}{\bibinfo{person}{Suhang Wang}, \bibinfo{person}{Yilin Wang}, \bibinfo{person}{Jiliang Tang}, \bibinfo{person}{Kai Shu}, \bibinfo{person}{Suhas Ranganath}, {and} \bibinfo{person}{Huan Liu}.} \bibinfo{year}{2017}\natexlab{b}.
\newblock \showarticletitle{What Your Images Reveal: Exploiting Visual Contents for Point-of-Interest Recommendation}. In \bibinfo{booktitle}{\emph{WWW}}. \bibinfo{pages}{391–400}.
\newblock


\bibitem[Wang et~al\mbox{.}(2018b)]%
        {SPMF}
\bibfield{author}{\bibinfo{person}{Weiqing Wang}, \bibinfo{person}{Hongzhi Yin}, \bibinfo{person}{Zi Huang}, \bibinfo{person}{Qinyong Wang}, \bibinfo{person}{Xingzhong Du}, {and} \bibinfo{person}{Quoc Viet~Hung Nguyen}.} \bibinfo{year}{2018}\natexlab{b}.
\newblock \showarticletitle{Streaming Ranking Based Recommender Systems}. In \bibinfo{booktitle}{\emph{SIGIR}}. \bibinfo{pages}{525–534}.
\newblock


\bibitem[Wang et~al\mbox{.}(2019)]%
        {wang2019neural}
\bibfield{author}{\bibinfo{person}{Xiang Wang}, \bibinfo{person}{Xiangnan He}, \bibinfo{person}{Meng Wang}, \bibinfo{person}{Fuli Feng}, {and} \bibinfo{person}{Tat-Seng Chua}.} \bibinfo{year}{2019}\natexlab{}.
\newblock \showarticletitle{Neural graph collaborative filtering}. In \bibinfo{booktitle}{\emph{SIGIR}}. \bibinfo{pages}{165--174}.
\newblock


\bibitem[Wang et~al\mbox{.}(2022)]%
        {autofield}
\bibfield{author}{\bibinfo{person}{Yejing Wang}, \bibinfo{person}{Xiangyu Zhao}, \bibinfo{person}{Tong Xu}, {and} \bibinfo{person}{Xian Wu}.} \bibinfo{year}{2022}\natexlab{}.
\newblock \showarticletitle{AutoField: Automating Feature Selection in Deep Recommender Systems}. In \bibinfo{booktitle}{\emph{WWW}}. \bibinfo{pages}{1977–1986}.
\newblock


\bibitem[White et~al\mbox{.}(2021)]%
        {white2021how}
\bibfield{author}{\bibinfo{person}{Colin White}, \bibinfo{person}{Arber Zela}, \bibinfo{person}{Binxin Ru}, \bibinfo{person}{Yang Liu}, {and} \bibinfo{person}{Frank Hutter}.} \bibinfo{year}{2021}\natexlab{}.
\newblock \showarticletitle{How Powerful are Performance Predictors in Neural Architecture Search?}. In \bibinfo{booktitle}{\emph{Advances in Neural Information Processing Systems}}. \bibinfo{pages}{28454--28469}.
\newblock


\bibitem[Xia et~al\mbox{.}(2022)]%
        {xiaxin2022odn}
\bibfield{author}{\bibinfo{person}{Xin Xia}, \bibinfo{person}{Hongzhi Yin}, \bibinfo{person}{Junliang Yu}, \bibinfo{person}{Qinyong Wang}, \bibinfo{person}{Guandong Xu}, {and} \bibinfo{person}{Quoc Viet~Hung Nguyen}.} \bibinfo{year}{2022}\natexlab{}.
\newblock \showarticletitle{On-Device Next-Item Recommendation with Self-Supervised Knowledge Distillation}. In \bibinfo{booktitle}{\emph{SIGIR}}. \bibinfo{pages}{546–555}.
\newblock


\bibitem[Xia et~al\mbox{.}(2023)]%
        {xiaxin2023eod}
\bibfield{author}{\bibinfo{person}{Xin Xia}, \bibinfo{person}{Junliang Yu}, \bibinfo{person}{Qinyong Wang}, \bibinfo{person}{Chaoqun Yang}, \bibinfo{person}{Nguyen Quoc~Viet Hung}, {and} \bibinfo{person}{Hongzhi Yin}.} \bibinfo{year}{2023}\natexlab{}.
\newblock \showarticletitle{Efficient On-Device Session-Based Recommendation}.
\newblock \bibinfo{journal}{\emph{ACM Trans. Inf. Syst.}} \bibinfo{volume}{41}, \bibinfo{number}{4} (\bibinfo{year}{2023}), \bibinfo{numpages}{24}~pages.
\newblock


\bibitem[Yin et~al\mbox{.}(2015)]%
        {yin2015jmo}
\bibfield{author}{\bibinfo{person}{Hongzhi Yin}, \bibinfo{person}{Bin Cui}, \bibinfo{person}{Zi Huang}, \bibinfo{person}{Weiqing Wang}, \bibinfo{person}{Xian Wu}, {and} \bibinfo{person}{Xiaofang Zhou}.} \bibinfo{year}{2015}\natexlab{}.
\newblock \showarticletitle{Joint Modeling of Users' Interests and Mobility Patterns for Point-of-Interest Recommendation}. In \bibinfo{booktitle}{\emph{Proceedings of the 23rd ACM International Conference on Multimedia}}. \bibinfo{pages}{819–822}.
\newblock


\bibitem[Yin et~al\mbox{.}(2024)]%
        {yin2024ondevicerec}
\bibfield{author}{\bibinfo{person}{Hongzhi Yin}, \bibinfo{person}{Liang Qu}, \bibinfo{person}{Tong Chen}, \bibinfo{person}{Wei Yuan}, \bibinfo{person}{Ruiqi Zheng}, \bibinfo{person}{Jing Long}, \bibinfo{person}{Xin Xia}, \bibinfo{person}{Yuhui Shi}, {and} \bibinfo{person}{Chengqi Zhang}.} \bibinfo{year}{2024}\natexlab{}.
\newblock \bibinfo{title}{On-Device Recommender Systems: A Comprehensive Survey}.
\newblock
\newblock
\showeprint[arxiv]{2401.11441}


\bibitem[Yu et~al\mbox{.}(2022)]%
        {graphaug}
\bibfield{author}{\bibinfo{person}{Junliang Yu}, \bibinfo{person}{Hongzhi Yin}, \bibinfo{person}{Xin Xia}, \bibinfo{person}{Tong Chen}, \bibinfo{person}{Lizhen Cui}, {and} \bibinfo{person}{Quoc Viet~Hung Nguyen}.} \bibinfo{year}{2022}\natexlab{}.
\newblock \showarticletitle{Are Graph Augmentations Necessary? Simple Graph Contrastive Learning for Recommendation}. In \bibinfo{booktitle}{\emph{SIGIR}}. \bibinfo{pages}{1294–1303}.
\newblock


\bibitem[Zhang et~al\mbox{.}(2024a)]%
        {changshuozhang2024rlt}
\bibfield{author}{\bibinfo{person}{Changshuo Zhang}, \bibinfo{person}{Sirui Chen}, \bibinfo{person}{Xiao Zhang}, \bibinfo{person}{Sunhao Dai}, \bibinfo{person}{Weijie Yu}, {and} \bibinfo{person}{Jun Xu}.} \bibinfo{year}{2024}\natexlab{a}.
\newblock \showarticletitle{Reinforcing Long-Term Performance in Recommender Systems with User-Oriented Exploration Policy}. In \bibinfo{booktitle}{\emph{SIGIR}}. \bibinfo{pages}{1850–1860}.
\newblock


\bibitem[Zhang et~al\mbox{.}(2024b)]%
        {changshuozhang2024qagcf}
\bibfield{author}{\bibinfo{person}{Changshuo Zhang}, \bibinfo{person}{Teng Shi}, \bibinfo{person}{Xiao Zhang}, \bibinfo{person}{Yanping Zheng}, \bibinfo{person}{Ruobing Xie}, \bibinfo{person}{Qi Liu}, \bibinfo{person}{Jun Xu}, {and} \bibinfo{person}{Ji-Rong Wen}.} \bibinfo{year}{2024}\natexlab{b}.
\newblock \bibinfo{title}{QAGCF: Graph Collaborative Filtering for Q\&A Recommendation}.
\newblock
\newblock
\showeprint[arxiv]{2406.04828}


\bibitem[Zhang et~al\mbox{.}(2019a)]%
        {tsample}
\bibfield{author}{\bibinfo{person}{Lingling Zhang}, \bibinfo{person}{Hong Jiang}, \bibinfo{person}{Fang Wang}, \bibinfo{person}{Dan Feng}, {and} \bibinfo{person}{Yanwen Xie}.} \bibinfo{year}{2019}\natexlab{a}.
\newblock \showarticletitle{T-Sample: A Dual Reservoir-Based Sampling Method for Characterizing Large Graph Streams}. In \bibinfo{booktitle}{\emph{ICDE}}. \bibinfo{pages}{1674--1677}.
\newblock


\bibitem[Zhang et~al\mbox{.}(2019b)]%
        {zhang2019deep}
\bibfield{author}{\bibinfo{person}{Shuai Zhang}, \bibinfo{person}{Lina Yao}, \bibinfo{person}{Aixin Sun}, {and} \bibinfo{person}{Yi Tay}.} \bibinfo{year}{2019}\natexlab{b}.
\newblock \showarticletitle{Deep learning based recommender system: A survey and new perspectives}.
\newblock \bibinfo{journal}{\emph{ACM Computing Surveys (CSUR)}} \bibinfo{volume}{52}, \bibinfo{number}{1} (\bibinfo{year}{2019}), \bibinfo{pages}{1--38}.
\newblock


\bibitem[Zhao et~al\mbox{.}(2020)]%
        {autoemb}
\bibfield{author}{\bibinfo{person}{Xiangyu Zhao}, \bibinfo{person}{Chong Wang}, \bibinfo{person}{Ming Chen}, \bibinfo{person}{Xudong Zheng}, \bibinfo{person}{Xiaobing Liu}, {and} \bibinfo{person}{Jiliang Tang}.} \bibinfo{year}{2020}\natexlab{}.
\newblock \showarticletitle{AutoEmb: Automated Embedding Dimensionality Search in Streaming Recommendations}. In \bibinfo{booktitle}{\emph{ICDM}}. \bibinfo{pages}{896--905}.
\newblock


\bibitem[Zheng et~al\mbox{.}(2024)]%
        {zhengruiqi2024pee}
\bibfield{author}{\bibinfo{person}{Ruiqi Zheng}, \bibinfo{person}{Liang Qu}, \bibinfo{person}{Tong Chen}, \bibinfo{person}{Kai Zheng}, \bibinfo{person}{Yuhui Shi}, {and} \bibinfo{person}{Hongzhi Yin}.} \bibinfo{year}{2024}\natexlab{}.
\newblock \showarticletitle{Personalized Elastic Embedding Learning for On-Device Recommendation}.
\newblock \bibinfo{journal}{\emph{IEEE Transactions on Knowledge and Data Engineering}}  \bibinfo{volume}{36} (\bibinfo{year}{2024}), \bibinfo{pages}{3363--3375}.
\newblock


\bibitem[Zheng et~al\mbox{.}(2022)]%
        {automlrecsurvey}
\bibfield{author}{\bibinfo{person}{Ruiqi Zheng}, \bibinfo{person}{Liang Qu}, \bibinfo{person}{Bin Cui}, \bibinfo{person}{Yuhui Shi}, {and} \bibinfo{person}{Hongzhi Yin}.} \bibinfo{year}{2022}\natexlab{}.
\newblock \showarticletitle{AutoML for Deep Recommender Systems: A Survey}.
\newblock \bibinfo{journal}{\emph{ACM Transactions on Information Systems}} \bibinfo{number}{101} (\bibinfo{year}{2022}), \bibinfo{pages}{1--38}.
\newblock


\bibitem[Zhou et~al\mbox{.}(2018)]%
        {rececom2}
\bibfield{author}{\bibinfo{person}{Meizi Zhou}, \bibinfo{person}{Zhuoye Ding}, \bibinfo{person}{Jiliang Tang}, {and} \bibinfo{person}{Dawei Yin}.} \bibinfo{year}{2018}\natexlab{}.
\newblock \showarticletitle{Micro Behaviors: A New Perspective in E-Commerce Recommender Systems}. In \bibinfo{booktitle}{\emph{ICDM}}. \bibinfo{pages}{727–735}.
\newblock


\end{thebibliography}

\end{document}